\documentclass[11pt,a4paper]{articlegm}
\usepackage{pdfpages}
\usepackage{graphicx}
\usepackage{lscape}
\usepackage{graphics}%
\usepackage{subfigure}
\usepackage{epsfig}
\usepackage{rotating}

\sffamily
\pdfoutput=1

\renewcommand{\labelitemii}{$\triangleright$}
\newcommand{\bitem}{\begin{list}{\labelitemi}{\itemsep=1.5cm}}
\newcommand{\eitem}{\end{list}}
\newcommand{\bitemt}{\begin{list}{\labelitemii}{\itemsep=0.7cm}}
\newcommand{\eitemt}{\end{list}}
\newcommand{\bitemtt}{\begin{list}{\labelitemiii}{\itemsep=0.5cm}}
\newcommand{\eitemtt}{\end{list}}

\newcommand{\bm}[1]{\mbox{\boldmath $#1$}}

\newcommand{\bX}{\bm{X}}

\newcommand{\bXi}{\mbox{$\bX_i$}}
\newcommand{\bXpi}{\mbox{$\bX'_i$}}

\newcommand{\bXidj}{\mbox{$\bX_{i(j)}$}}
\newcommand{\bXpidj}{\mbox{$\bX'_{i(j)}$}}

\newcommand{\bZ}{\bm{Z}}

\newcommand{\bZi}{\mbox{$\bZ_i$}}
\newcommand{\bZpi}{\mbox{$\bZ'_{i}$}}

\newcommand{\balpha}{\bm{\alpha}}
\newcommand{\bbeta}{\bm{\beta}}
\newcommand{\bdeta}{\bm{\eta}}

\newcommand{\bH}{\bm{H}}
\newcommand{\bI}{\bm{I}}

\begin{document}

\sffamily

\begin{center}

{\Large\bfseries\sffamily
Statistical model for overdispersed count outcome with many zeros: an approach for direct marginal inference}

\vspace*{5mm}

{\large\sffamily\bfseries Samuel Iddi$^{1}$ and Kwabena Doku-Amponsah$^{1}$}

\vspace*{5mm}

{\sffamily $^{1}$ University of Ghana, Department of Statistics, P. O. Box LG115, Legon, Ghana}
\end{center}

\begin{abstract}
Marginalized models are in great demand by most researchers in the life sciences particularly in clinical trials, epidemiology, health-economics, surveys and many others since they allow generalization of inference to the entire population under study. For count data, standard procedures such as the Poisson regression and negative binomial model provide population average inference for model parameters. However, occurrence of excess zero counts and lack of independence in empirical data have necessitated their extension to accommodate these phenomena. These extensions, though useful, complicates interpretations of effects. For example, the zero-inflated Poisson model accounts for the presence of excess zeros but the parameter estimates do not have a direct marginal inferential ability as its base model, the Poisson model. Marginalizations due to the presence of excess zeros are underdeveloped though demand for such is interestingly high.  The aim of this paper is to develop a marginalized model for zero-inflated univariate count outcome in the presence of overdispersion. Emphasis is placed on methodological development, efficient estimation of model parameters, implementation and application to two empirical studies. A simulation study is performed to assess the performance of the model. Results from the analysis of two case studies indicated that the refined procedure performs significantly better than models which do not simultaneously correct for overdispersion and presence of excess zero counts in terms of likelihood comparisons and AIC values. The simulation studies also supported these findings. In addition, the proposed technique yielded small biases and mean square errors for model parameters. To ensure that the proposed method enjoys widespread use, it is implemented using the SAS NLMIXED procedure with minimal coding efforts.

\vspace*{3mm}

\end{abstract}

\renewcommand{\baselinestretch}{1.25}

\section{Introduction}
\vspace*{-2mm}
Studies involving count data are widespread. They can be found in contemporary research areas such as in clinical trials, epidemiology studies, health-economics, surveys and other experiments in biopharmaceutical and bioinformatics. When the response of interest is of count type, the Poisson regression, which is placed within the generalized linear modeling (GLM) framework (Nelder and Wedderburn 1972, McCullagh and Nelder 1989, Agresti 2002), is routinely used to model the effect of covariates on the observed counts. Its application can be found in several research fields.

The most efficient way to make reliable inferences from well designed and executed studies is to choice an appropriate statistical model which reflects not only the design of the study but also certain characteristics of the data. The Poisson regression, though popular, fails to address certain attributes of the data and key design features and has led to several extensions. In the presence of many zero counts, especially in studies that involve 'rare' events, the Poisson regression is far from optimal. The zero-inflated Poisson (ZIP) model has been proposed to model count data with excessive zeros. Related to the presence of excess zeros is the phenomenon called overdispersion. The Poisson distribution, a member of the exponential family of distributions, is noted for having a strict mean-variance relationship which is often inadequate to capture the variability inherent in empirical data. In other to allow for inflation of the variance of the outcome, the negative binomial (NB) model has been developed and applied in many studies. Underdispersion is well possible but rarely encountered. For underdispersed data, the generalized Poisson, or perhaps the hurdle model is used. A broad overview of models and estimation methods for overdispersed data can be found in Hinde and Demétrio (1998ab).  As can be expected, excess zeros and overdispersion do occur together in practice. The zero-inflated negative binomial (ZINB) model is routinely used to handle both simultaneously and has also been implemented in several studies (Sheu and Liang, 1987).

Despite the useful extensions made to improve the Poisson regression in the presence of many zeros and overdispersion, interpretation of model parameters are hampered. Precisely, the marginal interpretation of effects of explanatory variables on the response is lost. Instead, the parameters have a latent class interpretation. This is because the ZI models assume separate models for the process generating excess counts and the positive counts. The implication is that, different sets of parameters are associated with a subpopulation of {\em at-risk} or {\em susceptible} and a subpopulation of {\em not-at-risk} or {\em non-susceptible} groups and hence inference targeted at the entire population is difficult to obtain. An approach for obtaining marginal inference is therefore required for count data with excess zeros.

Heagerty (1999) introduced a technique that does not alter the marginal interpretation of model parameters when normal random effect are employed to correct for lack of independence in longitudinal binary outcome. This marginalized multilevel model (MMM) defines separately a marginal mean model and a conditional mean model and the two models are held together by a so-called connector function. Iddi and Molenberghs (2012; 2013) extended this marginalized model to accommodate for overdispersion (COMMM model) in the presence of subject-specific random effects. Lee {\em et al\/} (2011) also proposed an extension of the MMM to zero-inflated clustered count data using the hurdle model (Mullahy 1986). The form of marginalization considered by these authors is over so-called {\em subject-specific} random effects, used to handle association in longitudinal or clustered data. Long {\em et al\/} (2014) adapted these ideas and proposed a marginalized model that estimates overall exposure effects in the ZIP model for univariate count outcome. The marginalized zero-inflated model (MZIP) eliminates the latent class interpretation of regression coefficients in the traditional ZIP model and instead allows for exposure effect on the entire population under study. However, this model is not suited for univariate count data exhibiting overdispersion. Therefore, this paper aims to refine the MZIP model to handle marginalization in the presence of excess zeros and also encompass overdispersion, due to unobserved heterogeneity, that naturally occur with count data. The modeling framework envisaged takes into account these attributes of the data as well as permit population average inference of model quantities. This guarantees efficient estimation of model parameters and ensures proper statistical inferences are made leading to valid research conclusions for policy decisions and recommendations. Also, this will help solve interpretation and implementation challenges faced by many applied analyst.

The rest of the paper is organized in the following order. Section~\ref{introcases} is used to introduce two motivating datasets; these are analyzed in Section~\ref{analysiscases}. A review of existing methodology is provided and the refined technique presented in Section~\ref{sectionmethodology}. The maximum likelihood estimation strategy used for fitting the models is discussed in Section~\ref{sectionmethodology} is the topic for Section~\ref{estimation}. Simulation results are discussed in Section~\ref{simulations}. The paper conclude with final remarks in Section~\ref{conclusions}.

\section{Case Studies\label{introcases}}
\vspace*{-2mm}
The main purpose of this section is to present two case studies used to illustrate the proposed methodology and how it compares with existing methods. The data resulting from these studies exhibit both overdispersion and zero inflated counts which are attributes investigated by the proposed technique. These studies are described in turn.

\subsection{A Clinical Trial in Epileptic Patients \label{epilep}}
\vspace*{-2mm}
The data are from a randomized, double-blind, parallel group, multi-center study for the comparison of placebo with a new anti-epileptic drug (AED), in combination with one or two other AED's (Faught {\em et al\/} 1996). Patients were randomized after a 12-week stabilization period for the use of AED's, and during which the number of seizures were counted. After that run-in period, 45 patients were assigned to the placebo group, 44 to the new treatment. Patients were measured weekly and followed (double-blind) during 16 weeks; thereafter they entered a long-term open-extension study. Some patients were followed for up to 27 weeks. The outcome of interest is the number of epileptic seizures experienced during the latest week, i.e., since the last time the outcome was measured. The research question is whether or not the additional new treatment reduces the number of epileptic seizures.

\begin{figure}[t]
\begin{center}
\includegraphics[width=1.0\textwidth,natwidth=610,natheight=650]{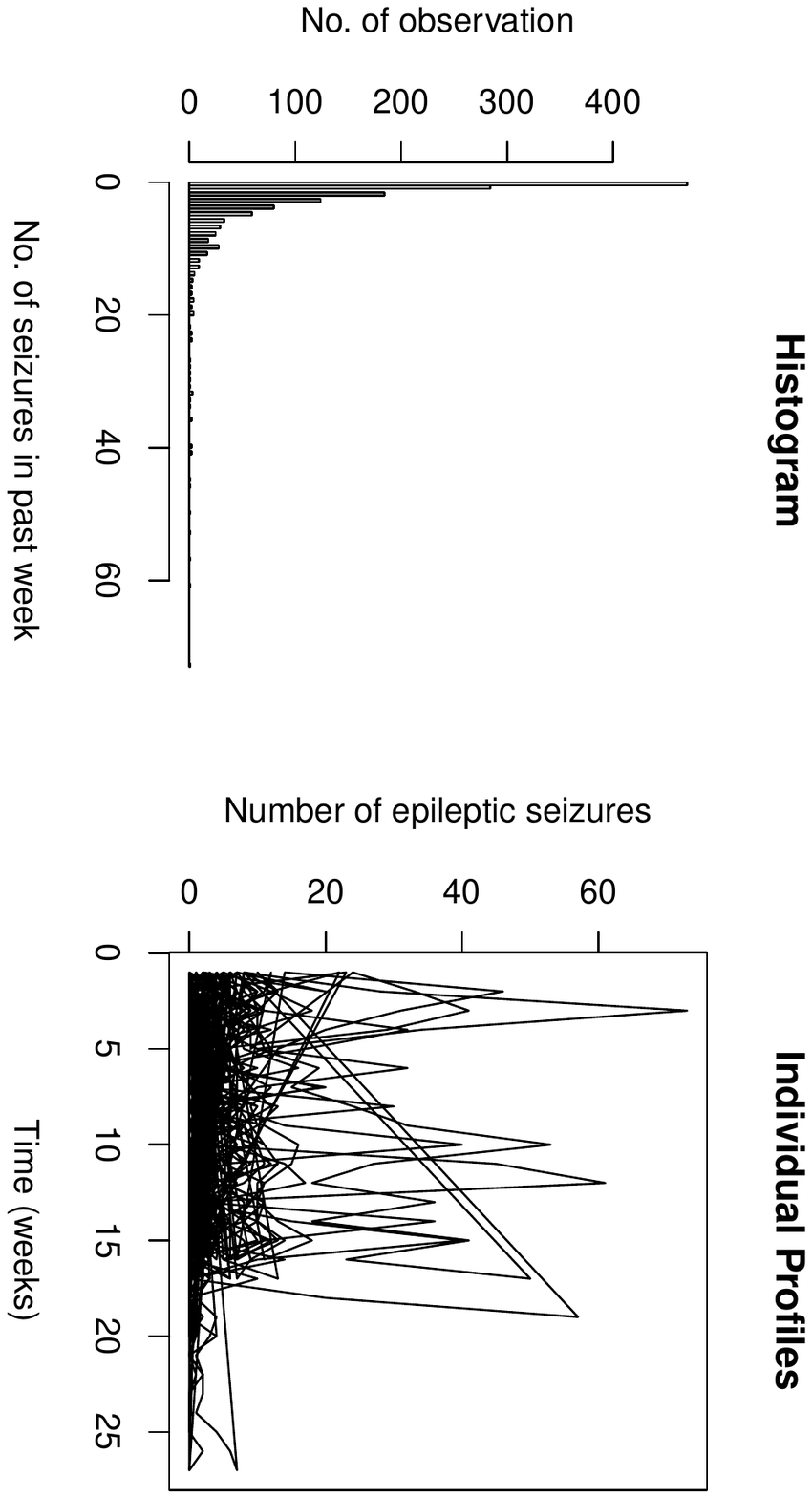}
\end{center}
\vspace*{-2.4in}
\caption{\em Epilepsy Data. Histogram and individual profiles.\label{epigraph}}
\end{figure}

In Figure~\ref{epigraph}, a histogram of the number of epileptic seizures shows a higher proportion of zero counts accounting for about 33\% of the data. Also, a simple descriptive statistics shows a very high variance of 37.70,  as compared to the empirical mean  of 3.18, an indication of overdispersion. This data is therefore suited for illustrating the proposed model.

\subsection{The Whitefly Study\label{jimma}}
\vspace*{-2mm}
This data resulted from a horticultural experiment design to examine the effect of six methods of applying insecticide imidacloprid to poinsettia plants. The data has previously been reported by van Iersel {\em et al\/} (2001) and analyzed in Hall and Zhang (2004). Using a randomized complete block design, treatment (method) was applied to 18 experimental units that consisted of a trio of 18 poinsettia plants (54 plants in total); repeated measurements were taken over 12 consecutive weeks. The experimental units were randomly assigned to the 6 treatments in 3 complete blocks. One of the study outcomes of interest here is the number of immature whiteflies after treatment out of a number of insects caged in one leaf per plant, prior to measurement of the response. The objective of the study was to investigate the best method to control silver-leaf whiteflies on the plants.

\begin{figure}[t]
\begin{center}
\includegraphics[width=1.0\textwidth,natwidth=610,natheight=650]{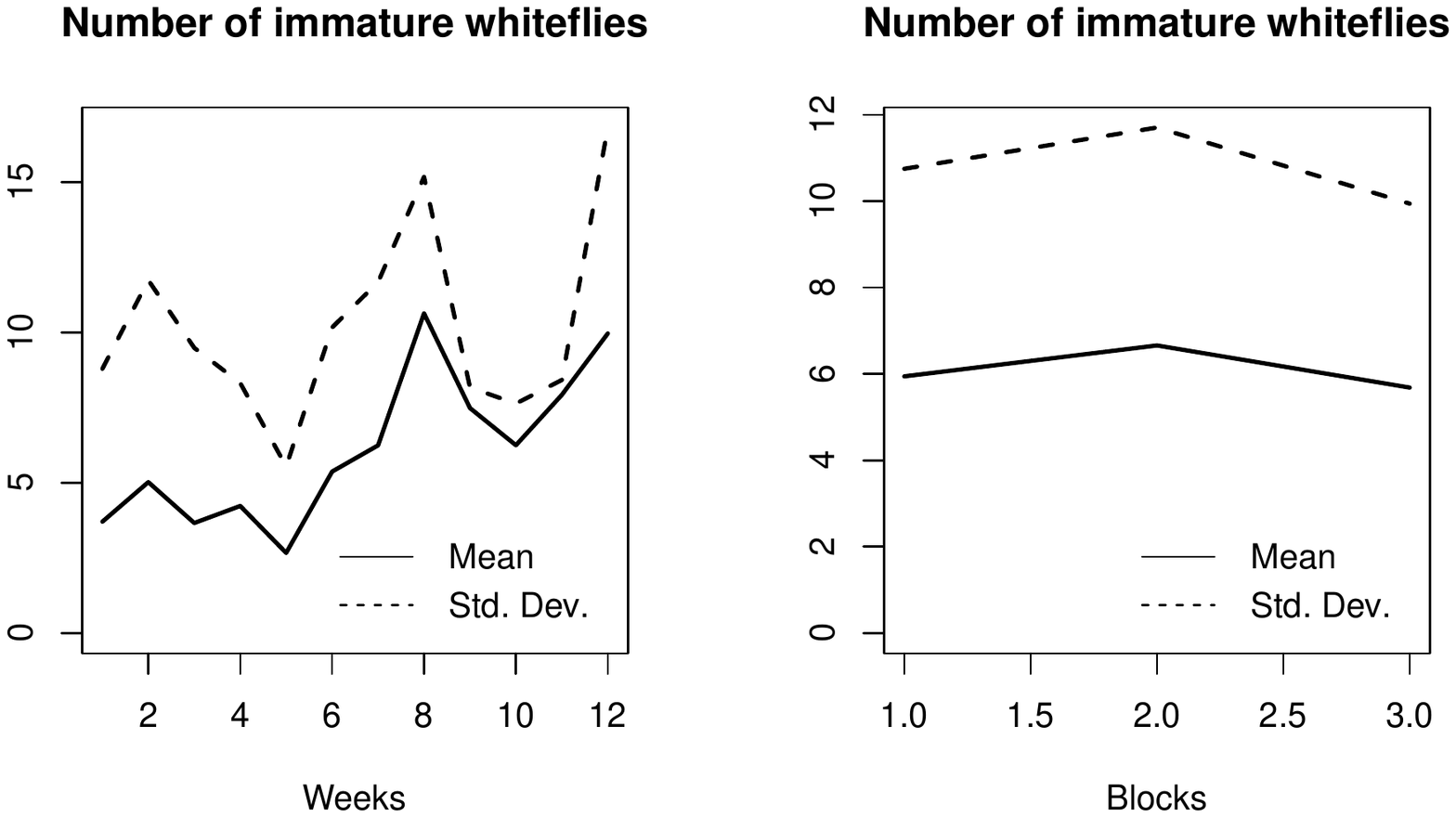}
\end{center}
\vspace*{-2.4in}
\caption{\em Whitefly Data. Means and standard deviations by time (panel 1) and block (panel 2).\label{wflygraph}}
\end{figure}

\begin{figure}[t]
\begin{center}
\includegraphics[width=0.75\textwidth,natwidth=610,natheight=650]{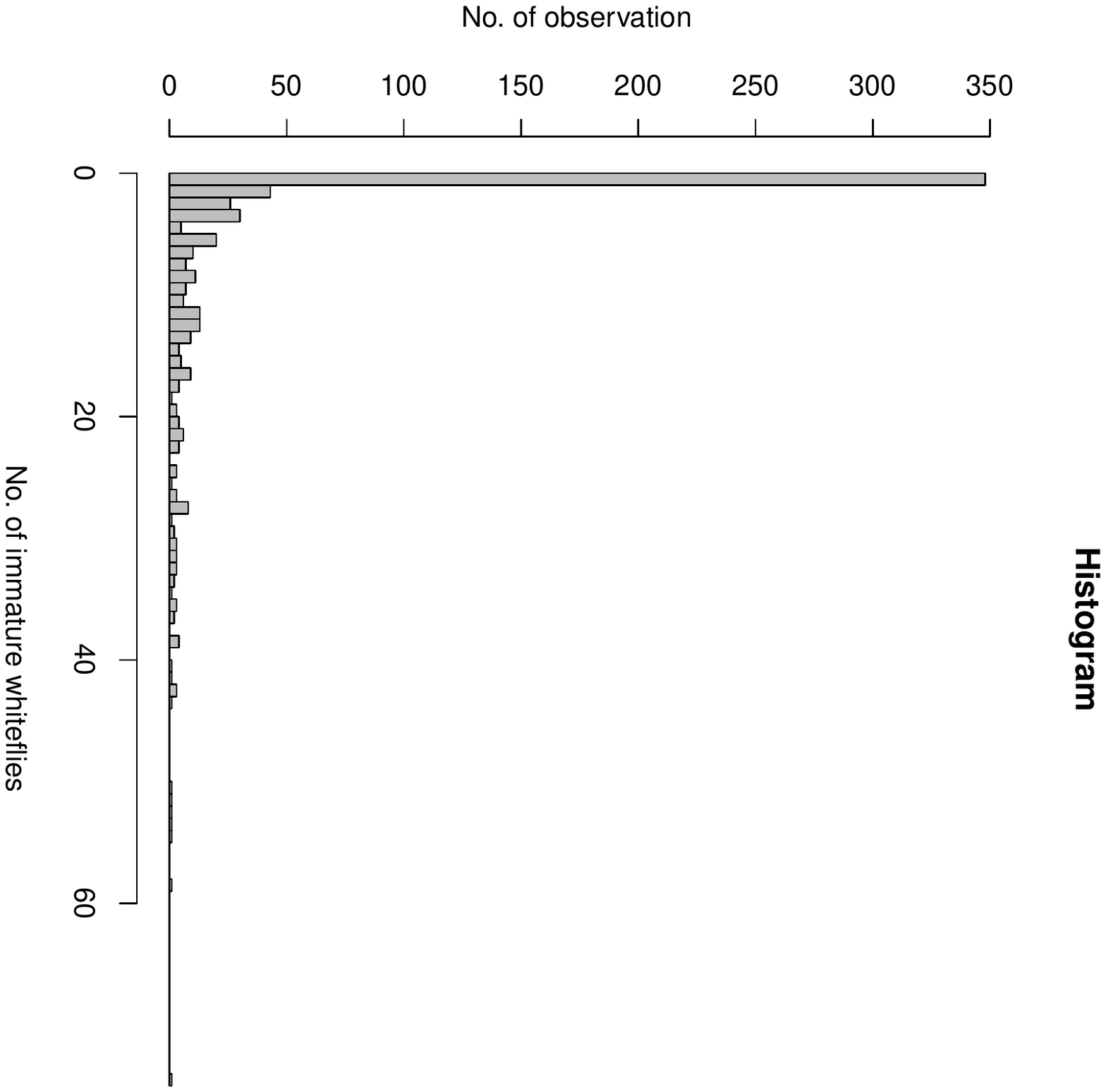}
\end{center}
\vspace*{-0.4in}
\caption{\em Whitefly Data. Histogram of the number of immature whiteflies.\label{wflyhist}}
\end{figure}

Figure~\ref{wflygraph} shows that, at every level of treatment and block, the variance is always higher than the mean, reflecting overdispersion. In Figure~\ref{wflyhist}, the histogram reveals higher occurrences of zero immature whiteflies, which cannot be accounted for by the variance function of a Poisson or negative binomial distribution.  It therefore seems sensible to apply the proposed technique.

\section{Methodology\label{sectionmethodology}}
\vspace*{-2mm}
\subsection{Notation}
\vspace*{-2mm}
Let $Y_i$ be the number of counts for an independent subject $i$ ($i=1,2,\dots,n$). Assume that together with the response, a set of regressors are recorded for each subject denoted $x_{ij}$ for $j=1,2,\dots,p$, where $p$ is the number of explanatory variables. Another set of covariates is represented as $z_{ik}$ which is a subset of $x_{ij}$ and $k\le j$. In vector notation, the covariates are written as $\bXi$ and $\bZi$ for the $i$th individual. Also, let the marginal and conditional expected count be denoted by $\mu_i$ and $\lambda_i$ respectively.

\subsection{Background Methodology\label{commm}}
\vspace{-2mm}
For independent count outcomes, the commonly used technique to evaluate the effect of explanatory variables on the response is the traditional Poisson regression. The response variable $Y_i$ is assumed to follow the Poisson distribution with mean $\mu_i$. The marginal mean is regressed on a set of covariate $\bXi$ using a log-link. Thus,
$$Y_i\sim \mbox{Poisson}(\mu_i)$$ and $$\mbox{log}(\mu_i)=\bXpi\bbeta$$
where $\bbeta$ is a vector of parameters associated with the vector of covariates, $\bXi$. The relationship between the response and the set of predictors is thus captured by $\bbeta$.

The Poisson regression model assumes, in its simplest form, that the marginal mean and variance of the response are equal. This strong assumption, often not tenable for empirical data due to heterogeneity introduced in the data when important covariates are omitted from the study, is relaxed by applying an overdispersed model.  A commonly used overdispersed model is the negative binomial regression model. It assumes that the counts follows a Poisson distribution with conditional mean $\lambda_i$. This mean is also allowed to follow the gamma distribution with shape and scale parameter $a$ and $b$ respectively. The resulting marginal distribution is the negative binomial distribution with density represented by
$$f(y_i)=\frac{\Gamma(a+y_i)}{\Gamma(a)y_i!}\left(\frac{b}{1+b}\right)^{y_i}\left(\frac{1}{1+b}\right)^a.$$
The first two marginal moments, using iterated expectations, are given respectively by
$$E(Y_i)=E\{E(Y_i|\lambda_i)\} =ab=\mu_i $$
$$\mbox{Var}(Y_i)=\mbox{Var}\{E(Y_i|\lambda_i)\}+E\{\mbox{Var}(Y_i|\lambda_i)\}=\mu_i(1+k\mu_i).$$
The parameter $k=\frac{1}{a}$ is called the overdispersed parameter. When $k=0$, the model reduces to the Poisson model. Since $k>0$, the model only models overdispersion and hence this model cannot be used to model underdispersion.
The NB regression models relates observed predictors to observed counts by taking $\mbox{log}(\mu_i)=X'_i\bbeta$.

Next, the Poisson and NB models assume that zero and non-zero counts are generated from the same mechanism. However, in the presence of excessive amount of zero counts, which occur mostly for rare events, these models are not optimal. This is because, they are unable to accommodate for the extra dispersion due to the presence of zeros. The zero-inflated Poisson (ZIP) has been proposed to address this issue. The model assumes that counts are rather generated by two processes. The first process generates zero counts with probability $\pi_i$ while the non-zero counts follows the Poisson distribution with parameter $\lambda_i$ and are realized with probability $(1-\pi)$. In addition, the model assumes that zero counts are generated from two sources based on the probabilities of the two processes. Thus,

$$Y_i \sim
\left\{
\begin{array}{cl}
0&\mbox{with probability}\quad\pi+(1-\pi)e^{-\lambda_i},\\
y_i&\mbox{with probability}\quad(1-\pi_i)e^{-\lambda_i}\frac{\lambda_i^{y_i}}{y_i!}, \quad y_i\in Z^{+}.
\end{array}
\right.
$$
The first two moments for ZIP model are given by
$$E(Y_i)=(1-\pi_i)\lambda_i=\mu_i$$
$$\mbox{Var}(Y_i)=\lambda_i(1-\pi_i)(1+\pi_i\lambda_i).$$
This reduces to the Poisson model when $\pi_i=0$. Note that the variance depends on the probability of zeros, $\pi_i$ and as $\pi_i$ approaches 1, the variance increases and thus accommodates greater dispersion in the data.
To fit this model, the logistic regression model is used to model the probability $\pi_i$ of zero counts and the log-linear $\mbox{Poisson}(\lambda_i)$ model for the positive realizations. For vector of covariates $\bZi$ and $\bXi$ with their associated vector of parameters $\balpha$ and $\bbeta$ respectively, the model specifications are as follows:
$$\mbox{logit}(\pi_i)=\bXpi\balpha$$ and $$\mbox{log}(\lambda_i)=\bXpi\bbeta.$$

\subsection{Marginal Effects and Incidence Density Ratio}
Marginal effect allows us to generalize the effects of predictors on the response variable to the entire population under consideration. Such inferences are based on the parameters associated with the predictors. In the traditional Poisson or negative binomial models, the regression coefficients are interpreted in terms of the differences in the logs of the expected counts for  a unit change in the predictor variables or as the log of the ratio of expected counts. Equivalently, the models are interpreted in terms of incident density (rate) ratio (IDR), obtained by exponentiating the regression estimates. Let $\mu_{i,j}$ and  $x_{i,j}$ be the mean and $j$th predictor variable evaluated at $j$ respectively. Also, let $\bXidj$ be a vector of predictors where the $j$th variable has been removed from $\bXi$ with associated vector of regression coefficients $\bbeta_{(j)}$. Then the IDR, the ratio of the marginal expected mean for a unit increase in the predictor variable $x_{i,j}$, is given by
$$\frac{E(Y_i|x_{i,j+1}=j+1)}{E(Y_i|x_{i,j}=j)}=\frac{\mu_{i,j+1}}{\mu_{i,j}}=\mbox{exp}({\beta_j})$$
where $\beta_j$ is the parameter associated with the $j$th predictor. This ratio is thus constant over the various levels of all other predictors in the regression model.

For the zero-inflated models, the marginal mean $\mu_i$ is of the form:
$$\mu_i=E(Y)=\frac{\mbox{exp}(\bXpi\bbeta)}{1+\mbox{exp}(\bZpi\balpha)}.$$
This depends on parameters associated with predictors in both components of the zero-inflated model.
Assume that the same predictors $\bXi$ are used in both the logistic and log-linear part of the models, then the IDR is expressed as
$$\frac{\mu_{i,j+1}}{\mu_{i,j}}=\mbox{exp}({\beta_j})\frac{1+\mbox{exp}\left({j\alpha_j + \bXpidj\balpha_{(j)}}\right)}{1+\mbox{exp}\left({(j+1)\alpha_j + \bXpidj\balpha_{(j)}}\right)}.$$
Unlike the Poisson and NB models, the IDR varies across the various levels of the predictors in the logistic part of the zero-inflated model. Only when $\alpha_j=0$ is the IDR constant across the levels of the extraneous predictors. Thus one has to employ a summary measure to obtain a single measure for IDR for a given predictor in the presence of the other predictors. Next, estimates of the variability of the IDR are obtained using the delta method or bootstrap resampling techniques. However, implementation of these techniques are cumbersome and require additional computational efforts since they are not readily available in standard software packages.

Recent development by Long {\em et al\/} (2014) allows analyst to fit a zero-inflated model with marginal effect of explanatory variables on the expected counts. The model also admits constant IDR for a given covariate in the presence of other predictors in both the logistic and the other component of the model. This model is reviewed in the next section and an extension to this procedure is proposed to accommodate for overdispersion.

\subsection{Proposed Methodology\label{zicommm}}
\vspace*{-2mm}
Long {\em et al\/} (2014) proposed an easy alternative to estimate overall exposure effects in a zero-inflated Poisson model. Instead of relating the mean of the process generating the positive counts, or the Poisson mean, $\lambda_i$ to predictors using the log-link, they expressed the marginal mean, $\mu_i$ in terms of predictors. The detailed model specifications are as follows:
$\mbox{logit}(\pi_i)=\bZpi\balpha$, $\mbox{log}(\lambda_i)=\delta_i$ and $\mbox{log}(\mu_i)=\bXpi\bbeta$ where $\delta_i$ is unknown function to be determined from $$\mu_i=(1-\pi_i)\lambda_i.$$ After substituting the various expressions and solving for $\delta_i$, we obtain $$\delta_i=\bXpi\bbeta+\mbox{log}(1+\mbox{exp}(\bZpi\balpha)).$$
The likelihood function is then modified based on these new expressions and is presented in Section~\ref{estimation}.

A marginal zero-inflated negative binomial model (MZINB), an extension of the MZIP model, is carried out to estimating marginal effect predictors on the marginal response. An added advantage to this useful extension is that, the model is able to correct for overdispersion due to the presence of both inflated zero counts and heterogeneity due to the absence of important predictors in the model. The latter is not addressed by the MZIP model.

The MZINB is also based on the zero-inflated negative binomial (ZINB) model. Let
$$Y_i \sim
\left\{
\begin{array}{cl}
0&\mbox{with probability}\quad\pi+(1-\pi)p^{\frac{1}{k}},\\
y_i&\mbox{with probability}\quad(1-\pi_i)\frac{\Gamma(y_i+\frac{1}{k})}{\Gamma(\frac{1}{k})y_i!}(1-p)^{y_i}p^{\frac{1}{k}}, \quad y_i\in Z^{+}.
\end{array}
\right.
$$
where $p=\frac{1}{1+k\lambda_i}.$ The marginal mean $\mu_i$ is similar to the mean from the ZIP model. However, the variance, which depends on the overdispersed parameter $k$ and $\pi_i$, takes the form
$$\mbox{Var}(Y_i)=\lambda_i(1-\pi_i)(1+\lambda_i(k+\pi_i)).$$
Thus, the model flexibly accounts for overdispersion due to the presence of excess zeros and heterogeneity due to the absence of omitted important explanatory variables.

To fit the MZINB model, we take $\mu_i=\mbox{exp}(\bXpi\bbeta)$ as opposed to $\lambda_i=\mbox{exp}(\bXpi\bbeta)$ in the traditional ZINB. Also, the mean of the positive counts or the negative binomial mean takes the form $\lambda_i=\mbox{exp}(\delta)$. The expression for $\delta_i$ is similarly to that of the MZIP model. The difference between the two procedures is clearer in their likelihood specification as discussed in Section~\ref{estimation}.

\section{Estimation \label{estimation}}
\vspace*{-2mm}
Several estimation routes, such as pseudo-likelihood (Aerts {\em et al\/}, 2002; Molenberghs and Vebeke, 2005), generalized estimating equations (Zeger, Liang, and Albert, 1988), and Bayesian methodology, are possible to in order to estimate the parameters in the models. In this paper, parameters in the models are estimated following maximum likelihood estimation technique. This estimation procedure, like many others, obtains a set of parameters that maximizes the marginal likelihood of the data. The likelihood for the marginal zero-inflated Poisson (MZIP) model is written as:
\begin{eqnarray}
L(\pi_i,\lambda_i|y_i)&=&\prod_{i=1}^{n}I(y_i=0)(1-\pi_i)\left[\frac{\pi_i}{1-\pi_i}+e^{-\lambda_i}\right]\prod_{i=1}^{n}I(y_i>0)(1-\pi_i)e^{-\lambda_i}\frac{\lambda_i^{y_i}}{y_i!} \label{llMZIP}\end{eqnarray}
Substituting $\pi_i=\mbox{expit}(\bZpi\balpha)$ and $\lambda_i=\mbox{exp}\left(\delta_i\right)=\left(1+\mbox{exp}(\bZpi\balpha)\right)\mbox{exp}(\bXpi\bbeta)$ into (\ref{llMZIP}), the likelihood in terms of the parameters $\balpha$ and $\bbeta$ becomes
\begin{eqnarray}
L(\balpha,\bbeta|y_i)&=&\prod_{i=1}^{n}\left(1+e^{\bZpi\balpha}\right)^{-1}\left[\prod_{i=1}^{n}I(y_i=0)\left(e^{\bZpi\balpha}+e^{-\left(1+\mbox{exp}(\bZpi\balpha)\right)\mbox{exp}\left(\bXpi\bbeta\right)}\right)\right.\nonumber\\
                     &\times&\left.\prod_{i=1}^{n}I(y_i=0)e^{-\left(1+\mbox{exp}({\bZpi\balpha})\right)\mbox{exp}\left({\bXpi\bbeta}\right)}\left(1+e^{\bZpi\balpha}\right)^{y_i}\frac{e^{\left(\bXpi\bbeta\right)y_i}}{y_i!}\right]\nonumber
\label{llMZIPm}
\end{eqnarray}

For the extended version (MZINB) with overdispersed parameter $k$, the likelihood is given by
\begin{eqnarray}
L(\pi_i,\lambda_i|y_i)&=&\prod_{i=1}^{n}I(y_i=0)(1-\pi_i)\left[\frac{\pi_i}{1-\pi_i}+\left(\frac{1}{1+k\lambda_i}\right)^{\frac{1}{k}}\right]\nonumber\\
&\times&\prod_{i=1}^{n}I(y_i>0)(1-\pi_i)\frac{\Gamma(y_i+\frac{1}{k})}{\Gamma(\frac{1}{k})y_i!}\left(1-\frac{1}{1+k\lambda_i}\right)^{y_i}\left(\frac{1}{1+k\lambda_i}\right)^{\frac{1}{k}} \label{llMZINB}\end{eqnarray}
Substituting expressions for $\pi_i$ and $\lambda_i$ into (\ref{llMZINB}) yields
\begin{eqnarray}
L(\balpha,\bbeta|y_i)&=&\prod_{i=1}^{n}\left(1+e^{\bZpi\balpha}\right)^{-1}\left[\prod_{i=1}^{n}I(y_i=0)\left(e^{\bZpi\balpha}+p_i^{\frac{1}{k}}\right)\right.\nonumber\\
                     &\times&\left.\prod_{i=1}^{n}I(y_i>0)\frac{\Gamma(y_i+\frac{1}{k})}{\Gamma(\frac{1}{k})y_i!}\left(1-p_i\right)^{y_i}p_i^{\frac{1}{k}}\right]\nonumber
\label{llMZINBm}
\end{eqnarray}
where $p_i=\frac{1}{1+k\left(1+\mbox{exp}(\bZpi\balpha)\right)\mbox{exp}(\bXpi\bbeta)}$
The maximum likelihood estimates $\hat{\balpha},\hat{\bbeta}$ and $\hat{k}$ are obtained through numerical maximization. The asymptotic variance-covariance matrix can be derived from the likelihood expression. We define the Hessian matrix of the mixed partial second derivatives of the log-likelihood, $l$ by $$\bH=\frac{\partial^2}{\partial \bdeta_i\partial \bdeta_j}l(\bdeta)$$ where $\bdeta=(\balpha,\bbeta,k)$. The Fisher's information matrix is given by $$\bI(\bdeta)=-E(\bH(\bdeta)).$$

The estimates can be obtained rather easily using the SAS software package procedure NLMIXED. The procedure allows to specify user defined log-likelihood and it returns, in addition, standard errors of the parameters. The standard errors are produced by taking the square root of the inverse of the Fisher's information matrix. Since the procedure performs all the numerical details, the applied analyst can avoid deriving close form score equations and the Fisher's information matrix.

The fit of the models are assessed using $-2\mbox{Log-likelihood}$ and the Akaike Information Criterion (AIC; Akaike, 1974). The model with the minimum value for each of the criteria is often considered the referred or 'best' model. AIC is calculated using the formula $\mbox{AIC}=-2\mbox{Log-likelihood}+2p$ where $p$ is the number of parameters in the model.

\section{Analysis of Case Studies\label{analysiscases}}
\vspace*{-2mm}
\subsection{Analysis of the Epilepsy Data\label{epilepdata}}
\vspace*{-2mm}
Six models are fitted to the data to their compare results. For each of the models, the dependent variable, $Y_i$, is the number of epileptic seizures experienced by patient $i$ which follows either a Poisson or negative binomial distribution. Treatment and time were treated as independent variables in the count part of the models and only time in the logistic part of the zero-inflated models. The Poisson, NB, MZIP and MZINB model can be viewed as 'marginal' models because they relates the marginal mean $\mu_i$ to the independent variables while in the ZIP and ZINB, the mean of the distribution of the positive counts, $\lambda_i$ are regressed on the predictors. Thus, in the Poisson, NB, MZIP and MZINB models,
$$\mbox{log}(\mu_i)=\beta_0+\beta_1\mbox{Treatment}_i+\beta_2\mbox{Time}_i.$$
For the zero-inflated models, the logistic regression model is specified as:
$$\mbox{logit}(\pi_i)=\alpha_0+\alpha_1\mbox{Time}_i.$$

The results of these models, parameter estimates and standard errors, are presented in Table~\ref{epi}. Generally, all parameters in the logistic-part of the zero-inflated models and the overdispersed parameter of the negative binomial models were found to be significant. Except for the ZINB model, Time was found to be significant in the count-part of the models. Treatment was found not to be significant in the Poisson and NB but significant for the MZIP model. However, the improved MZINB which acknowledged overdispersion, resulted in a non-significant results as the Poisson and NB model. For the ZIP and ZINB which have latent class interpretations, both Treatment and Time were significant for the former and not significant for the later. Results of model selection criteria, log-likelihood and AIC, varies for the different models. For the marginal models, the proposed MZINB model yielded the highest likelihood and smallest AIC value. Therefore, the proposed model seems to perform better than the rest of the marginal model and hence is essential for making inference and predictions.

\subsection{Analysis of the Whitefly Data\label{jimmadata}}
\vspace*{-2mm}
The outcome of this case study is the number of immature whiteflies, $Y_{ijk}$ for $i$th treatment in the $j$th block measured at the $k$th week and the independent variables are Treatment, Block and Week. The marginal mean model is given by:
$$\mbox{log}(\mu_{ijk})=\beta_0+\mbox{Block}_j+\mbox{Treatment}_i+\beta\mbox{Week}_k.$$ For the ZIP and ZINB models, $\lambda_{ijk}$ rather than $\mu_{ijk}$ is related to predictors. The probability of zero counts is modeled by:
$$\mbox{logit}(\pi_i)=\alpha_0+\alpha_1\mbox{Week}_k.$$
Week is treated as continuous and the other terms represent factor effects. Parameter estimates and standard errors of the fitted models are presented in Table~\ref{wfly}. All treatment levels and the effect of week were found to be significant in all the models. In addition, all parameters in the zero-inflated parts were also significant. The effect of Block1 and Block2 were not significant in all the negative binomial models whereas only in Block2 do we find a significant effect in the other models. Here again, in terms of inference, the proposed model results in slightly different parameter estimates and standard errors from the MZIP model although most of the parameters were significant, none of the block effect was significant in the broader model which properly accounted for overdispersion and excess zero counts.

It is observed from the model selection criteria that, extending the MZIP model by allowing overdispersion improved the model fit significantly (smallest AIC and highest likelihood). This again highlights the importance of acknowledging overdispersion in the model and hence can lead to better inference about the effect of independent variables on the response.

\section{Simulation Study\label{simulations}}
\vspace*{-2mm}
Simulations are carried out in this section to study some properties of the proposed methods and how it compares with the Poisson, negative binomial, and the marginal zero-inflated Poisson models. Large datasets were generated from the MZIP and MZINB models under difference settings. These are examined in turn.

\subsection{Data generated the from marginal zero-inflated model \label{simMZIP}}
\begin{figure}[htbp]
\centering
\includegraphics[width=11cm,height=7cm]{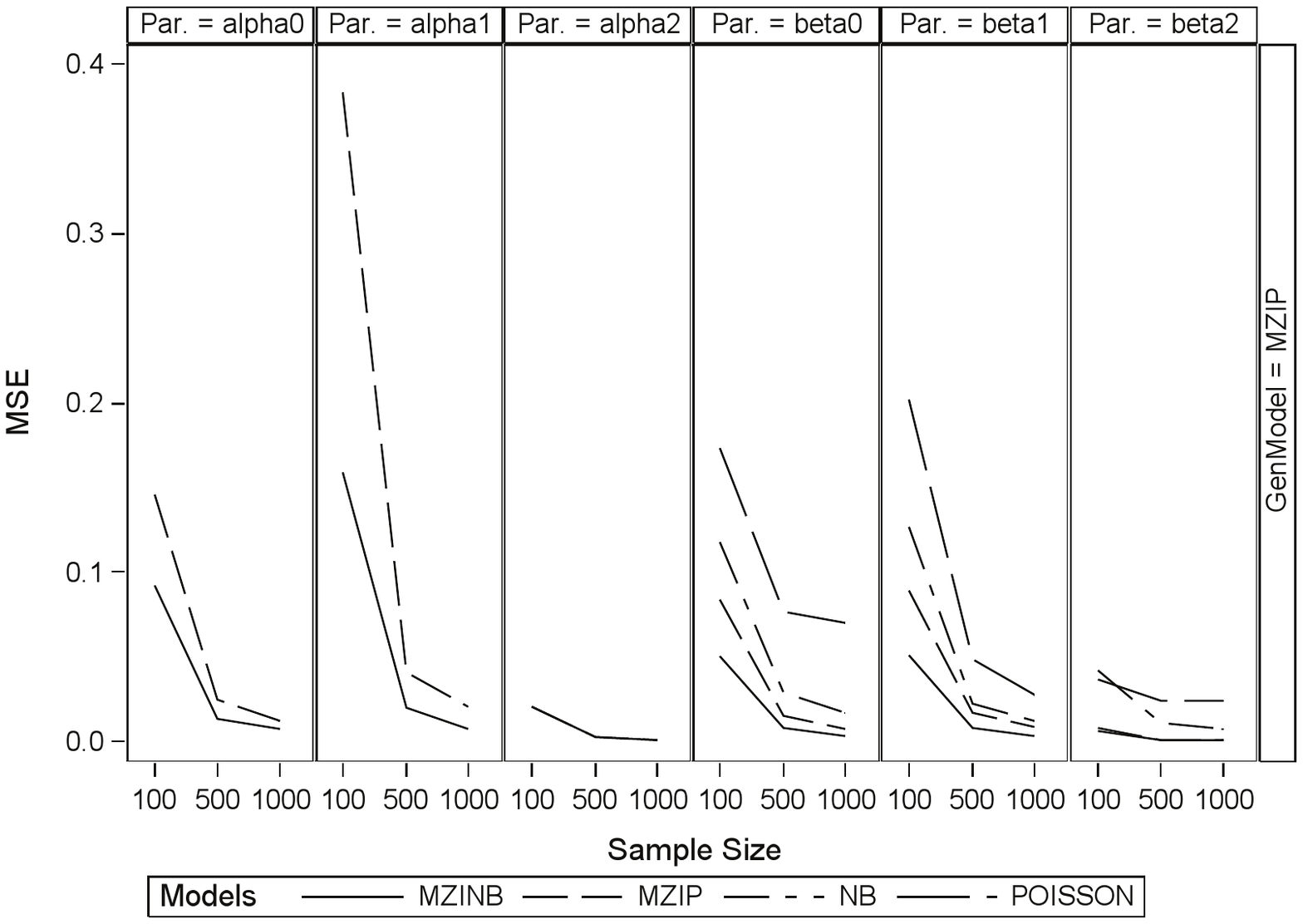}
\caption[]{\emph{Plot of MSE against sample size for all models (Data generated from MZIP)}}\label{mseMZIP}
\end{figure}

\begin{figure}[htbp]
\centering
\includegraphics[width=11cm,height=7cm]{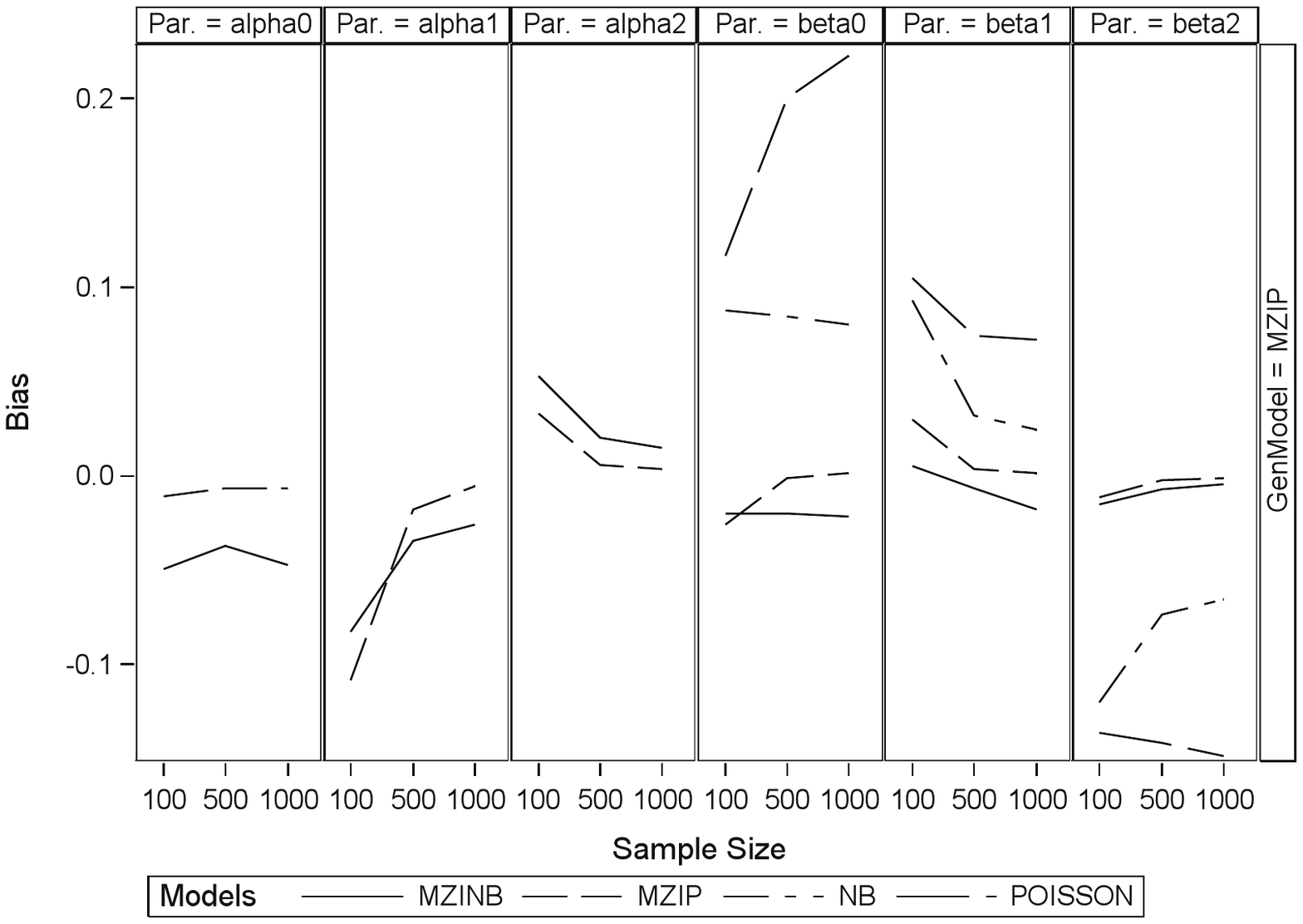}
\caption[]{\emph{Plot of bias against sample size for all models (Data generated from MZIP)}}\label{biasMZIP}
\end{figure}

\vspace*{-2mm}
This part of the simulation study is aimed at investigating the performance of the models particularly MZINB when data are only overdispersed due to the presence of excess zero counts. The simulated model utilized the following models:
$$\mbox{logit}(\pi_i)=\alpha_0+\alpha_1x_{i1}+\alpha_2x_{i2}$$
$$\mbox{log}(\mu_i)=\beta_0+\beta_1x_{i1}+\beta_2x_{i2}$$
where $i=1,…,n$, $x_{i1}\sim\mbox{Bernoulli}(0.5)$ and $x_{i2}$ follows a standard lognormal distribution. Zero-inflated counts were generated with the set of true parameters $$(\alpha_0,\alpha_1,\alpha_2,\beta_0,\beta_1,\beta_2) =(0.6,-2, 0.25, 0.25, 0.4,0.25).$$ For each sample size $n=(100,500,1000)$, 2000 datasets were generated from the marginal ZIP model and analyzed using the four models. Summary quantities, mean, standard errors, simulation based standard errors, bias, relative biased and mean square errors (MSE), are reported in Table~\ref{genMZIP1} and Table~\ref{genMZIP2}.  Generally, increase in sample size reduces the bias and MSE of the parameter estimates in all four models. The Poisson regression, followed by the NB models are the worst performers since they yielded large bias and MSE as depicted in Figure~\ref{biasMZIP} and Figure~\ref{mseMZIP} respectively. The MZINB model is slight biased compared to the MZIP model but this is compensated by the increase in precision resulting in smaller MSE compared to that of the MZIP model. This is not surprising given that the data were generated from the MZIP model. However, the broader MZINB model is able to precisely estimate the parameters as it also addresses the inflation of zeros and thus produces smaller measure of the overall variability.

\subsection{Data generated the from marginal zero-inflated negative binomial model \label{simMZINB}}
\begin{figure}[htbp]
\centering
\includegraphics[width=13cm,height=10cm]{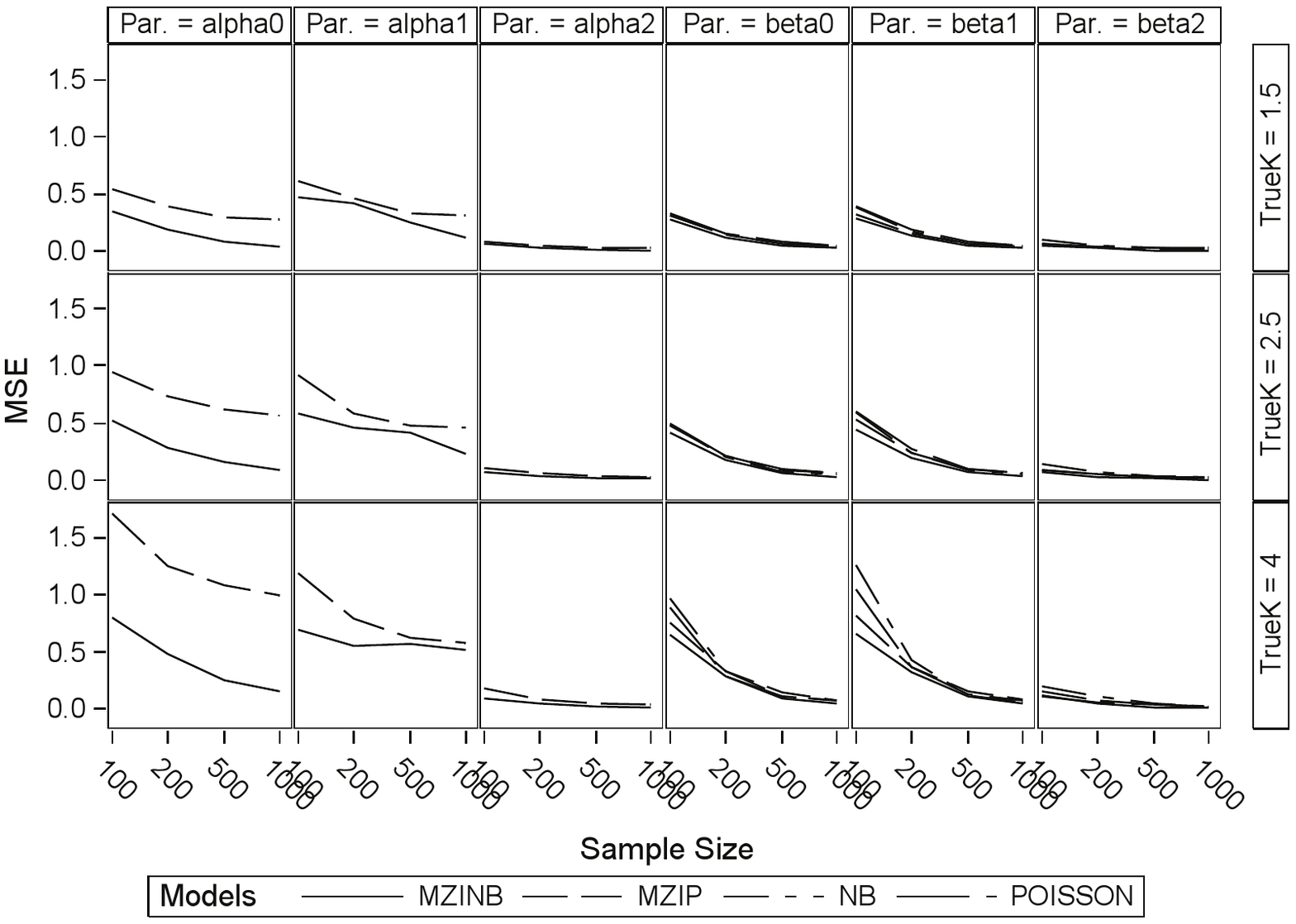}
\caption[]{\emph{Plot of MSE against sample size for all models (Data generated from MZINB)}}\label{mseMZINB}
\end{figure}

\begin{figure}[htbp]
\centering
\includegraphics[width=13cm,height=10cm]{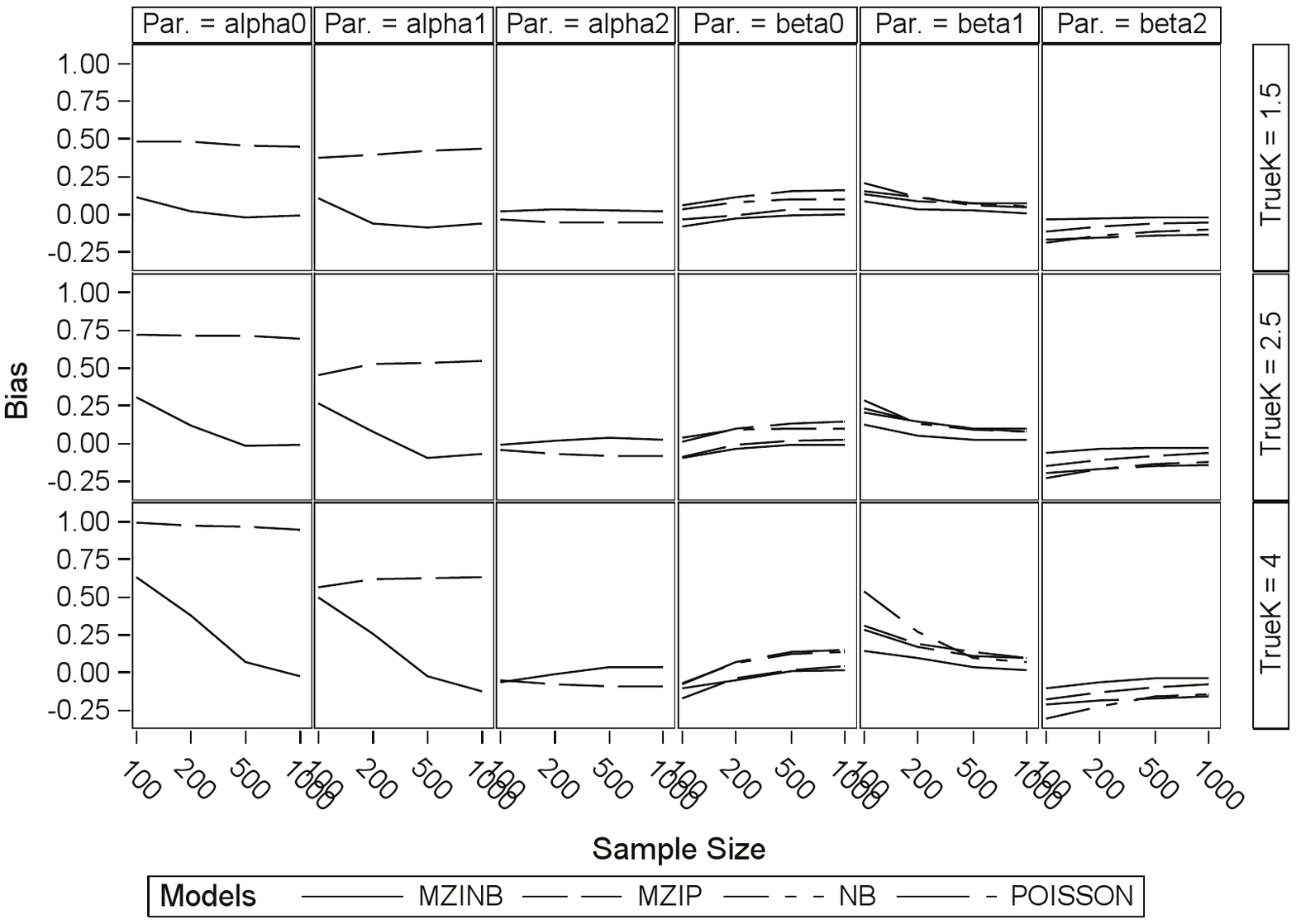}
\caption[]{\emph{Plot of bias against sample size for all models (Data generated from MZINB)}}\label{biasMZINB}
\end{figure}

\vspace*{-2mm}
The predictors used in this set of simulation study are similar to those used in Section~\ref{simMZIP}. To generated data from the marginal ZINB model, an additional parameter $k$ is required. The true parameter values are slightly modified, $$\alpha_0,\alpha_1,\alpha_2,\beta_0,\beta_1,\beta_2=(0.6,-2, 0.3, 0.25, 0.5,0.2).$$ The impact of different levels of overdispersion are assessed using different values of the overdispersed parameter $k=1.5, 2.5, 4$. For each value of $k$, 2000 simulated datasets were generated from the marginal ZINB model for different sample sizes, $n=(100, 200, 500, 1000)$ and each of the four models fitted. Simulation results are presented in Table~\ref{genMZINBBk1}, Table~\ref{genMZINBBk2}, Table~\ref{genMZINBBk3} for the Poison and negative binomial models, in Table~\ref{genMZINBMk1}, Table~\ref{genMZINBMk2}, Table~\ref{genMZINBMk3} for the MZIP and MZINB models, for the different values of $k$ respectively. Graph of bias and MSE against sample size are respectively depicted in Figure~\ref{biasMZINB} and Figure~\ref{mseMZINB}. From these results, it is observed that bias and MSE generally decreases with increasing sample size. Notably, the MSE increases with increase in overdispersed parameter but this diminishes with increase in sample size for all the models. The overdispersed parameter is poorly estimated by the NB model but better estimated by the proposed MZINB model with further improvements as sample size increases. Obviously, this is due to the excess zeros ignored by the negative binomial model but accounted for in the MZINB. Both the Poisson model and the negative binomial models fits poorly which is evident in the wide discrepancy between standard errors of parameters and the Monte Carlo based standard errors, large bias and MSE. The marginal ZINB model performs better than the rest of the models in terms of yielding the smallest bias as well as MSE for model parameters.

\section{Concluding Remarks\label{conclusions}}
\vspace*{-2mm}
It is commonly known that the Poisson regression is overly restrictive because of its mean-variance relationship and the presence of extra dispersion due to excess zeros. The ZIP and ZINB models are useful extension but give different interpretations of model parameters than the base models, namely, they have latent class rather than marginal interpretation. This paper has proposed an extension to the marginal ZIP model to address overdispersion. It has been shown, through the analysis of two case studies, that the extended model help improves model fit significantly and can help in drawing valid inference. Through simulation studies, it has been shown that even when data are generated from the MZIP model, the MZINB model tends to yield small MSE and bias. The MZIP model does worse when data are highly overdispersed.

The proposed model, due to it generality, can also be used to test the adequacy of MZIP model, i.e. by comparing the MZIP to the MZINB, we can test whether or not it is sufficient to use the MZIP model. If the data do not exhibit overdispersion, i.e. $k=0$, then the variance of the MZINB model reduces to the MZIP model. The marginal zero-inflated models are not a replacement to the traditional ZIP and ZINB model as the choice between latent class and marginal models will depend on the research question. If inference is targeted at providing population average inference about the effect of a variable or treatment, then it is easier and safer to begin with the proposed technique.

It is worth noting that, the proposed methodology is applicable to univariate data. Extension to correlated data such as longitudinal, repeated measures or clustered data may be required. One needs to be careful again with interpretation of model parameters when random effects are introduced in the proposed technique to accommodate association inherent in the data. Such an introduction will result in {\em subject-specific} interpretation instead of population average interpretation. Special techniques will be needed to obtain marginal inference in the presence of subject-specific random effects.

\section*{Acknowledgment}
\vspace*{-2mm}

\section*{\sffamily References}
\vspace*{-2mm}
\begin{description}
\item
Aerts, M., Geys, H., Molenberghs, G., and Ryan, L. (2002).
{\em Topics in Modelling of Clustered Data}.  London: Chapman \& Hall.

\item
Agresti, A. (2002).
{\em Categorical Data Analysis.}   New York: John Wiley \& Sons.

\item
Akaike, H. (1994).
A new look at the statistical model identification. {\em IEEE Transactions on Automatic Control}, {\bfseries 19}, 716--723.

\item
Faught, E., Wilder, B.J., Ramsay, R.E., Reife, R.A, Kramer, L.D., Pledger, G.W., and Karim, R.M. (1996).
Topiramate placebo-controlled dose-ranging trial in refractory partial epilepsy using 200-, 400-, and 600-mg daily dosage.
{\em Neurology}, {\bfseries 46}, 1684--1690.

\item
Hall, D.B. and Zhang, Z. (2004).
Marginal models for zero inflated clustered data. {\em Statistical Modelling}, {\bfseries 4}, 161--180.

\item {Heagerty, P.J.} (1999). Marginally specified logistic-normal models for longitudinal binary data. {\em Biometrics}, {\bfseries 55}, 688--698.

\item
Hinde, J. and Dem\'etrio, C.G.B. (1998a) Overdispersion: Models and estimation. {\em Computational Statistics and Data Analysis}, {\bfseries 27}, 151--170.

\item
Hinde, J. and Dem\'etrio, C.G.B. (1998b) {\em Overdispersion: Models and Estimation.} S\~ao Paulo: XIII Sinape.

\item Iddi, S. and Molenberghs. G. (2012).
A combined overdispersed and marginalized multilevel model. {\em Computational Statistics and Data Analysis}, {\bfseries 56}, 1944--1951.

\item Iddi, S. and Molenberghs. G. (2013).
A marginalized model for zero-inflated, overdispersed and correlated count data. {\em Electronic Journal of Applied Statistical Analysis}, {\bfseries 6}, 149--165.

\item Lee, K., Joo, Y., Song, J.J., and Harper, D.W. (2011).
Analysis of zero-inflated clustered count data: a marginalized model approach. {\em Computational Statistics and Data Analysis}, {\bfseries 55}, 824--837.

\item Long, D. L., Preisser, J., Herring, A., and Golin, C. (2014).
 A marginalized zero-inflated regression model with overall exposure effects. {\em Statistics in Medicine}, {\bfseries 33}, 5151--5165.

\item
McCullagh, P. and Nelder, J.A. (1989)
{\em Generalized Linear Models}. London: Chapman \& Hall/CRC.

\item
Molenberghs, G. and Verbeke, G. (2005).
{\em Models for Discrete Longitudinal Data.}   New York: Springer.

\item
Mullahy, J. (1986).
Specification and testing of some modified count data models.
{\em Journal of Econometrics}, {\bfseries 33}, 341--365.

\item
Nelder, J.A. and Wedderburn, R.W.M. (1972).
Generalized linear models. {\em Journal of the Royal Statistical Society, Series B},
{\bfseries 135}, 370--384.

\item van Iersel, M., Oetting, R., and Hall, D. B. (2001). Imidicloprid applications by subirrigation for control of silverleaf whitefly on poinsettia. {\em Journal of Economic Entomology}. {\bfseries 94}, 666--672.

\item
Zeger, S.L., Liang, K.-Y., and Albert, P.S. (1988). Models for longitudinal data: a generalized estimating equation approach. {\em Biometrics}, {\bfseries 44}, 1049--1060.

\end{description}


\addtolength{\tabcolsep}{-1mm}

\clearpage

\begin{landscape}
\begin{table}
\begin{center}
\vspace*{-0.60in}
\caption{\em Epilepsy Trial. Parameter estimates (standard errors) for the models.\label{epi}}
\sffamily
\vspace{0.2cm}
{\scriptsize\begin{tabular}{lccccccc}
\hline\hline
 &&\multicolumn{1}{c}{Poisson}&\multicolumn{1}{c}{NB}&\multicolumn{1}{c}{ZIP} &\multicolumn{1}{c}{ZINB}
	&\multicolumn{1}{c}{MZIP}&\multicolumn{1}{c}{MZINB}\\
\cline{3-4}\cline{5-6}\cline{7-8}
Effect&Par.&	Estimate(s.e) &	Estimate(s.e) &Estimate(s.e) &Estimate(s.e) &Estimate(s.e)&Estimate(s.e)\\
\hline
&\multicolumn{7}{c}{Count Part}\\
\hline	
Intercept &$\beta_0$& 1.3581(0.0316)*& 2.6520(0.1307)*& 1.5191(0.0328)*& 1.3007(0.0883)*& 1.3086(0.0454)*& 1.3484(0.0878)*\\
Treatment &$\beta_1$& 0.0232(0.0300) & 0.0115(0.0791) & 0.1603(0.0305)*& 0.0339(0.0788) &0 .1585(0.0305)*& 0.0274(0.0789)\\
    Time  &$\beta_2$&-0.0244(0.0029)*&-0.0249(0.0075)*&-0.0057(0.0030)*&-0.0146(0.0083) &-0.0263(0.0046)*&-0.0233(0.0077)*\\
\hline
&\multicolumn{7}{c}{Zero-Inflated Part}\\
\hline
Intercept&$\alpha_0$& 		   &                 &-1.7782(0.0815)*&-7.1223(1.3095)* &-1.3189(0.1180)* &-6.8819(1.2971)*\\
Time     &$\alpha_1$&              &             & 0.0394(0.0069)*& 0.2959(0.0644)* & 0.0627(0.0106)* & 0.2769(0.0604)* \\
\hline
Overdispersion&$k=\frac{1}{a}$&    &1.9002(0.0920)*&                  	&1.7851(0.1000)*&                 &1.7903(0.1017*)\\
-2Log-likelihood&$-2ll       $&11601&		6328.4&			9771.5&		6319.8& 9769.9           &6321.3\\
AIC              &            &11607&		6336.4&			9781.5&		6331.8& 9779.9           &6333.3\\
\hline \hline
\end{tabular}
}
\end{center}
\vspace*{-0.15in}
\hspace{1.20in}{\tiny $(^{*})$ Significant at $\alpha=0.05$}
\end{table}
\end{landscape}

\begin{landscape}
\begin{table}
\begin{center}
\vspace*{-0.60in}
\caption{\em Whitefly Data. Parameter estimates (standard errors) for the models. \label{wfly}}
\sffamily
\vspace{0.2cm}
{\scriptsize\begin{tabular}{lccccccc}
\hline\hline
 &&\multicolumn{1}{c}{Poisson}&\multicolumn{1}{c}{NB}&\multicolumn{1}{c}{ZIP} &\multicolumn{1}{c}{ZINB}
	&\multicolumn{1}{c}{MZIP}&\multicolumn{1}{c}{MZINB}\\
\cline{3-4}\cline{5-6}\cline{7-8}
Effect&Par.&	Estimate(s.e) &	Estimate(s.e) &Estimate(s.e) &Estimate(s.e) &Estimate(s.e)&Estimate(s.e)\\
\hline
&\multicolumn{7}{c}{Count Part}\\
\hline	
Intercept  &$\mu$  & 1.1405(0.0578)*& 0.5900(0.2709)*& 2.0333(0.0647)*& 1.8246(0.1920)*& 0.6441(0.1154)*& 0.5820(0.1974)*\\
    Block 1&       & 0.0270(0.0402) &-0.0099(0.1387) &-0.0305(0.0409) &-0.1222(0.1083) &-0.0301(0.0409) &-0.1292(0.1092) \\
    Block 2&       & 0.1535(0.0390)*& 0.0076(0.1375) & 0.0806(0.0397)*&-0.0702(0.1076) &0.08179(0.0397)*&-0.0703(0.1086) \\
Treatment 1&       &-1.0642(0.0762)*&-1.1141(0.1840)*&-0.8684(0.0779)*&-0.9978(0.1532)*&-0.8868(0.0780)*&-1.0373(0.1551)*\\
Treatment 2&       &-1.3630(0.0858)*&-1.2623(0.1895)*&-0.9228(0.0912)*&-1.1887(0.1733)*&-0.9165(0.0904)*&-1.1743(0.1784)*\\
Treatment 3&       &-2.0746(0.1169)*&-1.9587(0.2088)*&-1.5021(0.1686)*&-1.9426(0.1949)*&-1.4448(0.1620)*&-1.9351(0.2012)*\\
Treatment 4&       &-1.7587(0.1005)*&-1.7658(0.1967)*&-1.2744(0.1176)*&-1.6198(0.1806)*&-1.2724(0.1164)*&-1.6337(0.1855)*\\
Treatment 5&       & 1.3533(0.0431)*& 1.9507(0.1781)*& 0.9186(0.0441)*& 1.0263(0.1252)*& 0.9045(0.0442)*& 1.0082(0.1262)*\\
Week       &$\beta$& 0.0902(0.0048)*& 0.2288(0.0192)*& 0.0353(0.0052)*& 0.0660(0.0166)*& 0.1411(0.0107)*& 0.1768(0.0178)*\\
\hline
&\multicolumn{7}{c}{Zero-Inflated Part}\\
\hline
Intercept&$\alpha_0$& 		   &                 & 1.7012(0.2090)*& 1.6694(0.2716)*& 1.3894(0.1523)*& 1.3554(0.2246)*\\
Week     &$\alpha_1$&              &                 &-0.2824(0.0305)*&-0.3604(0.0482)*&-0.2312(0.0195)*&-0.3073(0.0394)*\\
\hline
Overdispersion&$k=\frac{1}{a}$&    &1.3587(0.1322)*  &                &0.4540 (0.0682)*&                &0.4577(0.0724)*\\
-2Log-likelihood&$-2ll       $&4174.6 &2692.7        &3278.3          &2619.5          &3279.7          &2626.6         \\
AIC              &            &4192.6 &2712.7        &3300.3          &2643.5          &3301.7          &2650.6         \\
\hline \hline
\end{tabular}
}
\end{center}
\vspace*{-0.15in}
\hspace{1.20in}{\tiny $(^{*})$ Significant at $\alpha=0.05$}
\end{table}
\end{landscape}

\clearpage

\begin{table}
\begin{center}
\vspace*{-0.60in}
\caption{\em Results of the Poisson and Negative Binomial Model based on 2000 Simulations from the MZIP.\label{genMZIP1}}
\sffamily
\vspace{0.2cm}
{\scriptsize\begin{tabular}{llrrrrrrrr}
\hline\hline
&&\multicolumn{3}{c}{Poisson}&&\multicolumn{3}{c}{Negative Binomial}\\
\cline{3-5}\cline{7-10}
\multicolumn{2}{l}{True parameters}&0.25&	0.4&	0.25&&0.25&	0.4&	0.25& -\\
\hline
Sample Size&Measure&$\beta_0$&$\beta_1$&$\beta_2$&&$\beta_0$&$\beta_1$&$\beta_2$&$k$\\
\hline\hline
100	&Estimate     &	0.3663&	0.5049&	 0.1133&&0.3375&0.4925&	0.1296	&2.5710\\
	&Std.~error   &	0.1235&	0.1371&	 0.0288&&0.3118&0.3588&	0.1148	&0.5725\\
	&SB~std.~err. &	0.4002&	0.4372&	 0.1336&&0.3319&0.3435&	0.1673	&0.6474\\
	&Bias	      &	0.1163&	0.1049&	-0.1367&&0.0875&0.0925&	-0.1204	&1.5710\\
	&Rel.~bias    &	0.4651&	0.2622&	-0.5467&&0.3500&0.2313&	-0.4816	&1.5710\\
	&MSE	      &	0.1737&	0.2021&	 0.0365&&0.1178&0.1265&	0.0425	&2.8872\\
\hline
500	&Estimate     &0.4496&	0.4744&	 0.1082&&0.3338	&0.4316&0.1765	&2.6976\\
	&Std.~error   &0.0501&	0.0583&	 0.0086&&0.1376	&0.1601&0.0473	&0.2608\\
	&SB~std.~err. &0.1904&	0.2066&	 0.0602&&0.1471	&0.1457&0.0736	&0.2869\\
	&Bias	      &0.1996&	0.0744&	-0.1418&&0.0838	&0.0316&-0.0735	&1.6976\\
	&Rel.~bias    &0.7985&	0.1860&	-0.5673&&0.3354	&0.0790&-0.2941	&1.6976\\
	&MSE	      &0.0761&	0.0482&	 0.0237&&0.0287	&0.0222&0.0108	&2.9642\\
\hline
1000&Estimate     &0.4727&	0.4718&	 0.1012&&0.3300&0.4238&	0.1839	&2.7214\\
	&Std.~error   &0.0347&	0.0409&	 0.0053&&0.0972&0.1132&	0.0329	&0.1852\\
	&SB~std.~err. &0.1417&	0.1515&	 0.0427&&0.1031&0.1070&	0.0488	&0.1981\\
	&Bias	      &0.2227&	0.0718&	-0.1488&&0.0800&0.0238&	-0.0661	&1.7214\\
	&Rel.~bias    &0.8906&	0.1795&	-0.5952&&0.3200&0.0594&	-0.2645	&1.7214\\
	&MSE	      &0.0697&	0.0281&	 0.0240&&0.0170&0.0120&	0.0068	&3.0025\\
\hline \hline
\end{tabular}
}
\end{center}
\end{table}

\begin{landscape}
\begin{table}
\begin{center}
\vspace*{-0.60in}
\caption{\em Results of the MZIP and MZINB based on 2000 Simulations from the MZIP.\label{genMZIP2}}
\sffamily
\vspace{0.2cm}
{\scriptsize\begin{tabular}{llrrrrrrrrrrrrrr}
\hline\hline
&&\multicolumn{6}{c}{Marginal ZIP}&&\multicolumn{6}{c}{Marginal ZINB}\\
\cline{3-8}\cline{10-16}
\multicolumn{2}{l}{True parameters}&0.25&	0.4&	0.25&0.6&-2&0.25&&0.25&	0.4& 0.25&0.6&-2&0.25& -\\
\hline
Sample Size&Measure&$\beta_0$&$\beta_1$&$\beta_2$&$\alpha_0$&$\alpha_1$&$\alpha_2$&&$\beta_0$&$\beta_1$&$\beta_2$&$\alpha_0$&$\alpha_1$&$\alpha_2$&$k$\\
\hline\hline
100	&Estimate     &0.2244	&0.4299& 0.2386	& 0.5890&-2.1089&0.2828&& 0.2297&0.4050&0.2351	&0.5503	&-2.0830&0.3027& 0.0263\\
	&Std.~error   &0.2807	&0.2908& 0.0851	& 0.3769&1.2940	&0.1315&& 0.2943&0.3003&0.1016	&0.4044	&0.5400	&0.1687& 0.0787\\
	&SB~std.~err. &0.2873	&0.2970& 0.0867	& 0.3814&0.6093	&0.1393&& 0.2231&0.2257&0.0796	&0.2998	&0.3903	&0.1343& 0.0295\\
	&Bias	      &-0.0256  &0.0299&-0.0114 &-0.0110&-0.1089&0.0328&&-0.0203&0.0050&-0.0149	&-0.0497&-0.0830&0.0527&-0.9737\\
	&Rel.~bias    &-0.1024  &0.0748&-0.0458 &-0.0184&0.0545	&0.1313&&-0.0813&0.0124&-0.0597	&-0.0829&0.0415	&0.2109&-0.9737\\
	&MSE	      &0.0832	&0.0891& 0.0076	& 0.1456&0.3831	&0.0205&& 0.0502&0.0510&0.0066	&0.0924	&0.1592	&0.0208& 0.9490\\
\hline
500	&Estimate     &0.2483	&0.4035& 0.2476& 0.5934&-2.0178&0.2556&& 0.2299&0.3931	&0.2427	&0.5625	&-2.0346&0.2695	& 0.0186\\
	&Std.~error   &0.1208	&0.1262& 0.0345& 0.1576& 0.2003&0.0463&& 0.1267&0.1309	&0.0403	&0.1697	&0.2275	&0.0598	& 0.0334\\
	&SB~std.~err. &0.1236	&0.1291& 0.0348& 0.1574& 0.2001&0.0472&& 0.0864&0.0908	&0.0309	&0.1090	&0.1360	&0.0446	& 0.0168\\
	&Bias	      &-0.0017	&0.0035&-0.0024&-0.0066&-0.0178&0.0056&&-0.0201&-0.0069&-0.0073&-0.0375&-0.0346&0.0195	&-0.9814\\
	&Rel.~bias    &-0.0067	&0.0088&-0.0094&-0.0110& 0.0089&0.0224&&-0.0805&-0.0172&-0.0290&-0.0626&0.0173	&0.0781	&-0.9814\\
	&MSE	      &0.0153	&0.0167& 0.0012& 0.0248& 0.0404&0.0023&& 0.0079&0.0083	&0.0010	&0.0133	&0.0197	&0.0024	& 0.9634\\
\hline
1000&Estimate     &0.2514&	0.4010&	 0.2481& 0.5937&-2.0062	&0.2534&& 0.2283&0.3816	&0.2455	&0.5524	&-2.0257&0.2648 & 0.0211\\
	&Std.~error   &0.0850&	0.0889&	 0.0240& 0.1093&0.1385	&0.0303&& 0.0895&0.0925	&0.0282	&0.1199	&0.1628	&0.0407 & 0.0253\\
	&SB~std.~err. &0.0865&	0.0928&	 0.0240& 0.1100&0.1433	&0.0305&& 0.0558&0.0522	&0.0216	&0.0671	&0.0815	&0.0290 & 0.0197\\
	&Bias	      &0.0014&	0.0010&	-0.0019&-0.0063&-0.0062	&0.0034&&-0.0217&-0.0184&-0.0045&-0.0476&-0.0257&0.0148 &-0.9789\\
	&Rel.~bias    &0.0055&	0.0026&	-0.0074&-0.0106&0.0031	&0.0135&&-0.0870&-0.0461&-0.0178&-0.0794&0.0129	&0.0591 &-0.9789\\
	&MSE	      &0.0075&	0.0086&	 0.0006& 0.0121&0.0206	&0.0009&& 0.0036&0.0031	&0.0005	&0.0068	&0.0073	&0.0011 & 0.9586\\
\hline \hline
\end{tabular}
}
\end{center}
\end{table}
\end{landscape}

\begin{table}
\begin{center}
\vspace*{-0.60in}
\caption{\em Results of the Poisson and Negative Binomial Model based on 2000 Simulations from the MZINB with $k=1.5$.\label{genMZINBBk1}}
\sffamily
\vspace{0.2cm}
{\scriptsize\begin{tabular}{llrrrrrrrr}
\hline\hline
&&\multicolumn{3}{c}{Poisson}&&\multicolumn{3}{c}{Negative Binomial}\\
\cline{3-5}\cline{7-10}
\multicolumn{2}{l}{True parameters}&0.25&	0.5&	0.2&&0.25&	0.5&	0.2&1.5\\
\hline
Sample Size&Measure&$\beta_0$&$\beta_1$&$\beta_2$&&$\beta_0$&$\beta_1$&$\beta_2$&$k$\\
\hline\hline
100	&Estimate     &0.3092&	0.6298&	 0.0294&&0.2857&	0.7024&	 0.0106&5.4097\\
	&Std.~error   &0.1378&	0.1509&	 0.0381&&0.4440&	0.5114&	 0.1787&1.2321\\
	&SB~std.~err. &0.5713&	0.6114&	 0.1779&&0.5488&	0.5846&	 0.2505&1.1858\\
	&Bias	      &0.0592&	0.1298&	-0.1706&&0.0357&	0.2024&	-0.1894&3.9097\\
	&Rel.~bias    &0.2369&	0.2597&	-0.8531&&0.1426&	0.4048&	-0.9471&2.6064\\
	&MSE	      &0.3299&	0.3907&	 0.0608&&0.3025&	0.3827&	 0.0986&16.6919\\
\hline
200	&Estimate     &0.3613&	0.5897&	 0.0486&&0.3283&	0.6104&	 0.0565&5.7296\\
	&Std.~error   &0.0905&	0.1011&	 0.0221&&0.3123&	0.3636&	 0.1183&0.9090\\
	&SB~std.~err. &0.3743&	0.4243&	 0.1114&&0.3609&	0.3878&	 0.1741&0.9412\\
	&Bias	      &0.1113&	0.0897&	-0.1514&&0.0783&	0.1104&	-0.1435&4.2296\\
	&Rel.~bias    &0.4451&	0.1793&	-0.7569&&0.3131&	0.2208&	-0.7174&2.8197\\
	&MSE	      &0.1525&	0.1881&	 0.0353&&0.1364&	0.1626&	 0.0509&18.7754\\
\hline
500	&Estimate     &0.3994&	0.5702&	 0.0564&&0.3456&	0.5693&	0.0832	&5.8368\\
	&Std.~error   &0.0544&	0.0620&	 0.0116&&0.1955&	0.2281&	0.0708	&0.5792\\
	&SB~std.~err. &0.2359&	0.2736&	 0.0639&&0.2238&	0.2413&	0.1060	&0.6369\\
	&Bias	      &0.1494&	0.0702&	-0.1436&&0.0956&	0.0693&	-0.1168	&4.3368\\
	&Rel.~bias    &0.5977&	0.1405&	-0.7182&&0.3825&	0.1385&	-0.5838	&2.8912\\
	&MSE	      &0.0780&	0.0798&	 0.0247&&0.0592&	0.0630&	0.0249 &19.2135\\
\hline
1000&Estimate     &0.4094&	0.5656&	0.0601&&0.3477&	0.5498&	 0.0979&5.8947\\
	&Std.~error   &0.0376&	0.0434&	0.0073&&0.1376&	0.1610&	 0.0490&0.4117\\
	&SB~std.~err. &0.1673&	0.1956&	0.0436&&0.1553&	0.1751&	 0.0735&0.4538\\
	&Bias	      &0.1594&	0.0656&-0.1399&&0.0977&	0.0498&	-0.1021&4.3947\\
	&Rel.~bias    &0.6377&	0.1313&-0.6996&&0.3908&	0.0997&	-0.5106&2.9298\\
	&MSE	      &0.0534&	0.0426&	0.0215&&0.0337&	0.0331&	 0.0158&19.5193\\
\hline \hline
\end{tabular}
}
\end{center}
\end{table}

\begin{table}
\begin{center}
\vspace*{-0.60in}
\caption{\em Results of the Poisson and Negative Binomial Model based on 2000 Simulations from the MZINB with $k=2.5$.\label{genMZINBBk2}}
\sffamily
\vspace{0.2cm}
{\scriptsize\begin{tabular}{llrrrrrrrr}
\hline\hline
&&\multicolumn{3}{c}{Poisson}&&\multicolumn{3}{c}{Negative Binomial}\\
\cline{3-5}\cline{7-10}
\multicolumn{2}{l}{True parameters}&0.25&	0.5&	0.2&&0.25&	0.5&	0.2&2.5 \\
\hline
Sample Size&Measure&$\beta_0$&$\beta_1$&$\beta_2$&&$\beta_0$&$\beta_1$&$\beta_2$&$k$\\
\hline\hline
100	&Estimate     &0.2575&	0.7050&	 0.0010&&0.2875&0.7876&	-0.0375&6.5425\\
	&Std.~error   &0.1463&	0.1587&	 0.0421&&0.4939&0.5651&	 0.2042&1.5199\\
	&SB~std.~err. &0.7074&	0.7483&	 0.2157&&0.6946&0.7131&	 0.2896&1.1279\\
	&Bias	      &0.0075&	0.2050&	-0.1990&&0.0375&0.2876&	-0.2375&4.0425\\
	&Rel.~bias    &0.0299&	0.4100&	-0.9952&&0.1501&0.5752&	-1.1876&1.6170\\
	&MSE	      &0.5005&	0.6020&	 0.0861&&0.4839&0.5912&	 0.1403&17.6140\\
\hline
200	&Estimate     &0.3430&	0.6371&	 0.0289&&0.3356&0.6353&	0.0268&	7.7645\\
	&Std.~error   &0.0934&	0.1035&	 0.0237&&0.3639&0.4211&	0.1396&	1.2892\\
	&SB~std.~err. &0.4526&	0.5034&	 0.1313&&0.4388&0.4694&	0.2003&	1.2649\\
	&Bias	      &0.0930&	0.1371&	-0.1711&&0.0856&0.1353&-0.1732&	5.2645\\
	&Rel.~bias    &0.3719&	0.2741&	-0.8555&&0.3423&0.2707&-0.8660&	2.1058\\
	&MSE	      &0.2135&	0.2722&	 0.0465&&0.1999&0.2386&	0.0701&	29.3149\\
\hline
500	&Estimate     &0.3855&	0.6003&	0.0454&&0.3461&	0.5912&	0.0616	&7.9611\\
	&Std.~error   &0.0556&	0.0628&	0.0124&&0.2269&	0.2646&	0.0835	&0.8285\\
	&SB~std.~err. &0.2683&	0.2998&	0.0718&&0.2534&	0.2833&	0.1194	&0.8805\\
	&Bias	      &0.1355&	0.1003&-0.1546&&0.0961&	0.0912&	-0.1384	&5.4611\\
	&Rel.~bias    &0.5419&	0.2006&-0.7729&&0.3845&	0.1825&	-0.6919	&2.1844\\
	&MSE	      &0.0903&	0.0999&	0.0291&&0.0734&	0.0886&	0.0334	&30.5989\\
\hline
1000&Estimate     &0.3916&	0.5978&	0.0526&&0.3491&	0.5738&	0.0765	&8.0249\\
	&Std.~error   &0.0384&	0.0439&	0.0079&&0.1597&	0.1865&	0.0579	&0.5880\\
	&SB~std.~err. &0.1971&	0.2193&	0.0470&&0.1812&	0.2011&	0.0826	&0.6328\\
	&Bias	      &0.1416&	0.0978&-0.1474&&0.0991&	0.0738&	-0.1235	&5.5249\\
	&Rel.~bias    &0.5664&	0.1957&-0.7369&&0.3963&	0.1476&	-0.6176	&2.2100\\
	&MSE	      &0.0589&	0.0577&	0.0239&&0.0427&	0.0459&	0.0221	&30.9250\\
\hline \hline
\end{tabular}
}
\end{center}
\end{table}

\begin{table}
\begin{center}
\vspace*{-0.60in}
\caption{\em Results of the Poisson and Negative Binomial Model based on 2000 Simulations from the MZINB with $k=4.0$.\label{genMZINBBk3}}
\sffamily
\vspace{0.2cm}
{\scriptsize\begin{tabular}{llrrrrrrrr}
\hline\hline
&&\multicolumn{3}{c}{Poisson}&&\multicolumn{3}{c}{Negative Binomial}\\
\cline{3-5}\cline{7-10}
\multicolumn{2}{l}{True parameters}&0.25&	0.5&	0.2&&0.25&	0.5&	0.2&4.0 \\
\hline
Sample Size&Measure&$\beta_0$&$\beta_1$&$\beta_2$&&$\beta_0$&$\beta_1$&$\beta_2$&$k$\\
\hline\hline
100	&Estimate     &0.1684	&0.7823&-0.0203&&0.1770	&1.0371&-0.1067	&7.1537\\
	&Std.~error   &0.1577	&0.1700& 0.0456&&0.5262	&0.5997&0.2233	&1.7016\\
	&SB~std.~err. &0.9353	&0.9822& 0.2331&&0.9783	&0.9851&0.3289	&1.0319\\
	&Bias	      &-0.0816	&0.2823&-0.2203&&-0.0730&0.5371&-0.3067	&3.1537\\
	&Rel.~bias    &-0.3262	&0.5646&-1.1015&&-0.2920&1.0743&-1.5337	&0.7884\\
	&MSE	      &0.8814	&1.0444& 0.1029&&0.9624	&1.2589&0.2022	&11.0106\\
\hline
200	&Estimate     &0.3179&	0.6659&	0.0082&&0.3142&	0.7655&	-0.0251	&9.5471\\
	&Std.~error   &0.0968&	0.1071&	0.0255&&0.4059&	0.4685&	0.1622	&1.6513\\
	&SB~std.~err. &0.5230&	0.5784&	0.1370&&0.5683&	0.5915&	0.2319	&1.2085\\
	&Bias	      &0.0679&	0.1659&-0.1918&&0.0642&	0.2655&	-0.2251	&5.5471\\
	&Rel.~bias    &0.2715&	0.3319&-0.9589&&0.2569&	0.5310&	-1.1257	&1.3868\\
	&MSE	      &0.2781&	0.3621&	0.0556&&0.3271&	0.4204&	0.1044	&32.2308\\
\hline
500	&Estimate     &0.3800&	0.6090&	0.0266&&0.3693&	0.5982&	0.0352	&11.0496\\
	&Std.~error   &0.0568&	0.0641&	0.0135&&0.2663&	0.3101&	0.1012	&1.2283\\
	&SB~std.~err. &0.3088&	0.3516&	0.0770&&0.3066&	0.3316&	0.1382	&1.3245\\
	&Bias	      &0.1300&	0.1090&-0.1734&&0.1193&	0.0982&	-0.1648	&7.0496\\
	&Rel.~bias    &0.5202&	0.2179&-0.8672&&0.4770&	0.1964&	-0.8241	&1.7624\\
	&MSE	      &0.1123&	0.1355&	0.0360&&0.1082&	0.1196&	0.0463	&51.4512\\
\hline
1000	&Estimate     &0.4016&	0.5994&	0.0339&&0.3802&	0.5726&	0.0530	&11.1191\\
	&Std.~error   &0.0390&	0.0444&	0.0086&&0.1867&	0.2181&	0.0692	&0.8688\\
	&SB~std.~err. &0.2124&	0.2429&	0.0501&&0.2219&	0.2387&	0.0926	&0.9441\\
	&Bias	      &0.1516&	0.0994&-0.1661&&0.1302&	0.0726&	-0.1470	&7.1191\\
	&Rel.~bias    &0.6062&	0.1988&-0.8305&&0.5209&	0.1452&	-0.7350	&1.7798\\
	&MSE	      &0.0681&	0.0689&	0.0301&&0.0662&	0.0622&	0.0302	&51.5729\\
\hline \hline
\end{tabular}
}
\end{center}
\end{table}

\begin{landscape}
\begin{table}
\begin{center}
\vspace*{-0.60in}
\caption{\em Results of the MZIP and MZINB based on 2000 Simulations from the MZINB with $k=1.5$.\label{genMZINBMk1}}
\sffamily
\vspace{0.2cm}
{\scriptsize\begin{tabular}{llrrrrrrrrrrrrrr}
\hline\hline
&&\multicolumn{6}{c}{Marginal ZIP}&&\multicolumn{6}{c}{Marginal ZINB}\\
\cline{3-8}\cline{10-16}
\multicolumn{2}{l}{True parameters}&0.25&	0.4&	0.25&0.6&-2&0.25&&0.25&	0.4& 0.25&0.6&-2&0.25&1.5 \\
\hline
Sample Size&Measure&$\beta_0$&$\beta_1$&$\beta_2$&$\alpha_0$&$\alpha_1$&$\alpha_2$&&$\beta_0$&$\beta_1$&$\beta_2$&$\alpha_0$&$\alpha_1$&$\alpha_2$&$k$\\
\hline\hline
100	&Estimate     &0.2081	&0.6462&0.0846	&1.0801&-1.6256	&0.2626&&0.1663	&0.5856&0.1579	&0.7147&-1.8969	&0.3120&1.0762\\
	&Std.~error   &0.3470	&0.3606&0.1137	&0.4191&0.4778	&0.1496&&0.4684	&0.4815&0.1856	&0.5773&0.7612	&0.2237&0.5769\\
	&SB~std.~err. &0.5502	&0.5504&0.2189	&0.5599&0.6912	&0.2821&&0.5143	&0.5270&0.2037	&0.5767&0.6795	&0.2342&0.5210\\
	&Bias	      &-0.0419	&0.1462&-0.1154	&0.4801&0.3744 &-0.0374&&-0.0837&0.0856&-0.0421	&0.1147&0.1031	&0.0120&-0.4238\\
	&Rel.~bias    &-0.1677	&0.2923&-0.5771	&0.8002&-0.1872&-0.1247&&-0.3347&0.1713&-0.2103	&0.1912&-0.0515	&0.0399&-0.2825\\
	&MSE	      &0.3045	&0.3243&0.0612	&0.5440&0.6179	&0.0810&&0.2715	&0.2851&0.0433	&0.3457&0.4723	&0.0550&0.4510\\
\hline
200	&Estimate     &0.2429	&0.6103&0.1145	&1.0780&-1.6090	&0.2451&&0.2252	&0.5378&0.1701	&0.6133&-2.0630	&0.3326&1.3548\\
	&Std.~error   &0.2372	&0.2480&0.0737	&0.2848&0.3270	&0.0945&&0.3334	&0.3461&0.1303	&0.4329&0.6513	&0.1627&0.5269\\
	&SB~std.~err. &0.3702	&0.3764&0.1450	&0.4046&0.5502	&0.2064&&0.3500	&0.3569&0.1384	&0.4303&0.6397	&0.1715&0.5144\\
	&Bias	      &-0.0071	&0.1103&-0.0855	&0.4780&0.3910 &-0.0549&&-0.0248&0.0378&-0.0299	&0.0133&-0.0630	&0.0326&-0.1452\\
	&Rel.~bias    &-0.0283	&0.2206&-0.4276	&0.7966&-0.1955&-0.1830&&-0.0993&0.0756&-0.1494	&0.0221&0.0315	&0.1086&-0.0968\\
	&MSE	      &0.1371	&0.1538&0.0283	&0.3922&0.4556	&0.0456&&0.1231	&0.1288&0.0200	&0.1853&0.4132	&0.0305&0.2857\\
\hline
500	&Estimate     &0.2835&	0.5575&	0.1323	&1.0569&-1.5780	&0.2452&&0.2388	&0.5230&0.1798	&0.5823	&-2.0920&0.3252&1.4916\\
	&Std.~error   &0.1466&	0.1541&	0.0454	&0.1752&0.2021	&0.0563&&0.2107	&0.2202&0.0807	&0.2813	&0.4426	&0.1017&0.3831\\
	&SB~std.~err. &0.2355&	0.2359&	0.0960	&0.2849&0.3940	&0.1548&&0.2103	&0.2190&0.0789	&0.2971	&0.4887	&0.1089&0.4073\\
	&Bias	      &0.0335&	0.0575&	-0.0677	&0.4569&0.4220 &-0.0548&&-0.0112&0.0230&-0.0202	&-0.0177&-0.0920&0.0252&-0.0084\\
	&Rel.~bias    &0.1341&	0.1149&	-0.3387	&0.7614&-0.2110&-0.1826&&-0.0447&0.0459&-0.1012	&-0.0295&0.0460	&0.0838&-0.0056\\
	&MSE	      &0.0566&	0.0590&	0.0138	&0.2899&0.3333	&0.0270&&0.0444	&0.0485&0.0066	&0.0886	&0.2473	&0.0125&0.1660\\
\hline
1000&Estimate     &0.2848&	0.5465&	0.1438	&1.0488&-1.5647	&0.2449&&0.2521&0.5105&	0.1831	&0.5853	&-2.0648&0.3142&1.5169\\
	&Std.~error   &0.1029&	0.1082&	0.0316	&0.1223&0.1412	&0.0382&&0.1486&0.1556&	0.0562	&0.1963	&0.2998	&0.0690&0.2778\\
	&SB~std.~err. &0.1819&	0.1730&	0.0797	&0.2555&0.3417	&0.1399&&0.1473&0.1574&	0.0558	&0.2050	&0.3453	&0.0723&0.3143\\
	&Bias	      &0.0348&	0.0465&	-0.0562	&0.4488&0.4353 &-0.0551&&0.0021&0.0105&	-0.0169	&-0.0147&-0.0648&0.0142&0.0169\\
	&Rel.~bias    &0.1394&	0.0929&	-0.2809	&0.7480&-0.2177&-0.1837&&0.0085&0.0209&	-0.0844	&-0.0245&0.0324	&0.0472&0.0112\\
	&MSE	      &0.0343&	0.0321&	0.0095	&0.2667&0.3062	&0.0226&&0.0217&0.0249&	0.0034	&0.0422	&0.1234	&0.0054&0.0991\\
\hline \hline
\end{tabular}
}
\end{center}
\end{table}
\end{landscape}

\begin{landscape}
\begin{table}
\begin{center}
\vspace*{-0.60in}
\caption{\em Results of the MZIP and MZINB based on 2000 Simulations from the MZINB with $k=2.5$.\label{genMZINBMk2}}
\sffamily
\vspace{0.2cm}
{\scriptsize\begin{tabular}{llrrrrrrrrrrrrrr}
\hline\hline
&&\multicolumn{6}{c}{Marginal ZIP}&&\multicolumn{6}{c}{Marginal ZINB}\\
\cline{3-8}\cline{10-16}
\multicolumn{2}{l}{True parameters}&0.25&	0.4&	0.25&0.6&-2&0.25&&0.25&	0.4& 0.25&0.6&-2&0.25&2.5 \\
\hline
Sample Size&Measure&$\beta_0$&$\beta_1$&$\beta_2$&$\alpha_0$&$\alpha_1$&$\alpha_2$&&$\beta_0$&$\beta_1$&$\beta_2$&$\alpha_0$&$\alpha_1$&$\alpha_2$&$k$\\
\hline\hline
100	&Estimate     &0.1600	&0.7298&0.0502	&1.3178&-1.5475	&0.2563&&0.1518	&0.6231&0.1310	&0.9047&-1.7356	&0.2858	& 1.4111\\
	&Std.~error   &0.3885	&0.4042&0.1277	&0.4524&0.5033	&0.1620&&0.5490	&0.5618&0.2216	&0.6362&0.7660	&0.2405	& 0.8202\\
	&SB~std.~err. &0.6822	&0.6926&0.2565	&0.6517&0.8456	&0.3243&&0.6394	&0.6495&0.2656	&0.6488&0.7164	&0.2694	& 0.6836\\
	&Bias	      &-0.0900	&0.2298&-0.1498	&0.7178&0.4525 &-0.0437&&-0.0982&0.1231&-0.0690	&0.3047&0.2644	&-0.0142&-1.0889\\
	&Rel.~bias    &-0.3601	&0.4596&-0.7492	&1.1963&-0.2262&-0.1456&&-0.3930&0.2462&-0.3451	&0.5079&-0.1322	&-0.0473&-0.4356\\
	&MSE	      &0.4735	&0.5325&0.0882	&0.9399&0.9198	&0.1071&&0.4185	&0.4370&0.0753	&0.5138&0.5831	&0.0728	& 1.6530\\
\hline
200	&Estimate     &0.2417	&0.6398&0.0900	&1.3144&-1.4714	&0.2235&&0.2103	&0.5544&0.1593	&0.7112&-1.9228	&0.3165&1.9562\\
	&Std.~error   &0.2612	&0.2737&0.0819	&0.3042&0.3415	&0.1014&&0.3915	&0.4050&0.1569	&0.5124&0.7038	&0.1774&0.8609\\
	&SB~std.~err. &0.4655	&0.4691&0.1833	&0.4668&0.5497	&0.2358&&0.4241	&0.4329&0.1683	&0.5250&0.6743	&0.2003&0.8099\\
	&Bias	      &-0.0083	&0.1398&-0.1100	&0.7144&0.5286 &-0.0765&&-0.0397&0.0544&-0.0407	&0.1112&0.0772	&0.0165&-0.5438\\
	&Rel.~bias    &-0.0332	&0.2797&-0.5501	&1.1907&-0.2643&-0.2551&&-0.1587&0.1088&-0.2033	&0.1853&-0.0386	&0.0548&-0.2175\\
	&MSE	      &0.2168	&0.2396&0.0457	&0.7283&0.5816	&0.0615&&0.1814	&0.1904&0.0300	&0.2880&0.4606	&0.0404&0.9517\\
\hline
500	&Estimate     &0.2678&	0.5995&	0.1205	&1.3126&-1.4671	&0.2142&&0.2432	&0.5235&0.1679	&0.5841	&-2.0984&0.3303&2.4125\\
	&Std.~error   &0.1613&	0.1699&	0.0490	&0.1872&0.2118	&0.0593&&0.2503	&0.2615&0.0992	&0.3661	&0.5779	&0.1209&0.7187\\
	&SB~std.~err. &0.2956&	0.2848&	0.1213	&0.3255&0.4278	&0.1801&&0.2478	&0.2627&0.0993	&0.3856	&0.6339	&0.1320&0.7563\\
	&Bias	      &0.0178&	0.0995&	-0.0795	&0.7126&0.5329 &-0.0858&&-0.0068&0.0235&-0.0321	&-0.0159&-0.0984&0.0303&-0.0875\\
	&Rel.~bias    &0.0714&	0.1989&	-0.3975	&1.1876&-0.2664&-0.2861&&-0.0271&0.0470&-0.1607	&-0.0265&0.0492	&0.1010&-0.0350\\
	&MSE	      &0.0877&	0.0910&	0.0210	&0.6137&0.4670	&0.0398&&0.0615	&0.0696&0.0109	&0.1489	&0.4115	&0.0183&0.5796\\
\hline
1000&Estimate     &0.2788&	0.5790&	0.1351	&1.2932&-1.4544	&0.2158&&0.2430	&0.5282&0.1726	&0.5914	&-2.0698&0.3229& 2.4703\\
	&Std.~error   &0.1125&	0.1191&	0.0339	&0.1301&0.1479	&0.0404&&0.1775	&0.1861&0.0695	&0.2657	&0.4043	&0.0846& 0.5516\\
	&SB~std.~err. &0.2312&	0.2196&	0.0937	&0.2843&0.3990	&0.1512&&0.1707	&0.1860&0.0674	&0.2835	&0.4649	&0.0918& 0.6032\\
	&Bias	      &0.0288&	0.0790&	-0.0649	&0.6932&0.5456 &-0.0842&&-0.0070&0.0282&-0.0274	&-0.0086&-0.0698&0.0229&-0.0297\\
	&Rel.~bias    &0.1152&	0.1580&	-0.3243	&1.1553&-0.2728&-0.2808&&-0.0279&0.0564&-0.1369	&-0.0143&0.0349	&0.0764&-0.0119\\
	&MSE	      &0.0543&	0.0545&	0.0130	&0.5614&0.4569	&0.0300&&0.0292	&0.0354&0.0053	&0.0804	&0.2210	&0.0090& 0.3647\\
\hline \hline
\end{tabular}
}
\end{center}
\end{table}
\end{landscape}

\begin{landscape}
\begin{table}
\begin{center}
\vspace*{-0.60in}
\caption{\em Results of the MZIP and MZINB based on 2000 Simulations from the MZINB with $k=4.0$.\label{genMZINBMk3}}
\sffamily
\vspace{0.2cm}
{\scriptsize\begin{tabular}{llrrrrrrrrrrrrrr}
\hline\hline
&&\multicolumn{6}{c}{Marginal ZIP}&&\multicolumn{6}{c}{Marginal ZINB}\\
\cline{3-8}\cline{10-16}
\multicolumn{2}{l}{True parameters}&0.25&	0.4&	0.25&0.6&-2&0.25&&0.25&	0.4& 0.25&0.6&-2&0.25&4.0 \\
\hline
Sample Size&Measure&$\beta_0$&$\beta_1$&$\beta_2$&$\alpha_0$&$\alpha_1$&$\alpha_2$&&$\beta_0$&$\beta_1$&$\beta_2$&$\alpha_0$&$\alpha_1$&$\alpha_2$&$k$\\
\hline\hline
100	&Estimate     &0.0788	&0.8071&0.0213	&1.5906&-1.4433	&0.2480&&0.1410	&0.6424&0.0911	&1.2338&-1.5003	&0.2327	&1.4905\\
	&Std.~error   &0.4449	&0.4622&0.1478	&0.5015&0.5456	&0.1799&&0.6294	&0.6440&0.2568	&0.6429&0.7035	&0.2485	&0.9487\\
	&SB~std.~err. &0.8530	&0.8487&0.3516	&0.8520&0.9348	&0.4156&&0.7957	&0.8026&0.3292	&0.6308&0.6706	&0.2991	&0.6639\\
	&Bias	      &-0.1712	&0.3071&-0.1787	&0.9906&0.5567 &-0.0520&&-0.1090&0.1424&-0.1089	&0.6338&0.4997	&-0.0673&-2.5095\\
	&Rel.~bias    &-0.6850	&0.6142&-0.8933	&1.6510&-0.2784&-0.1733&&-0.4361&0.2847&-0.5445	&1.0564&-0.2498	&-0.2242&-0.6274\\
	&MSE	      &0.7569	&0.8146&0.1556	&1.7072&1.1838	&0.1754&&0.6450	&0.6644&0.1202	&0.7996&0.6994	&0.0940	&6.7384\\
\hline
200	&Estimate     &0.2107	&0.6894&0.0610	&1.5762&-1.3851	&0.2175&&0.1927	&0.5966&0.1356	&0.9747&-1.7435	&0.2852	&2.3238\\
	&Std.~error   &0.2944	&0.3095&0.0938	&0.3319&0.3677	&0.1115&&0.4583	&0.4722&0.1857	&0.5390&0.6496	&0.1821	&1.1179\\
	&SB~std.~err. &0.5777	&0.5716&0.2302	&0.5376&0.6375	&0.2782&&0.5335	&0.5543&0.2068	&0.5834&0.6971	&0.2123	&1.0093\\
	&Bias	      &-0.0393	&0.1894&-0.1390	&0.9762&0.6149 &-0.0825&&-0.0573&0.0966&-0.0644	&0.3747&0.2565	&-0.0148&-1.6762\\
	&Rel.~bias    &-0.1572	&0.3788&-0.6949	&1.6270&-0.3074&-0.2749&&-0.2290&0.1933&-0.3221	&0.6245&-0.1283	&-0.0495&-0.4190\\
	&MSE	      &0.3353	&0.3626&0.0723	&1.2420&0.7845	&0.0842&&0.2879	&0.3166&0.0469	&0.4808&0.5517	&0.0453	&3.8283\\
\hline
500	&Estimate     &0.2677&	0.6281&	0.0988	&1.5610&-1.3802	&0.2080&&0.2529&0.5291&	0.1586	&0.6726&-2.0281	&0.3301&3.4567\\
	&Std.~error   &0.1799&	0.1903&	0.0550	&0.2027&0.2267	&0.0644&&0.2978&0.3107&	0.1219	&0.4395&0.6389	&0.1365&1.1602\\
	&SB~std.~err. &0.3812&	0.3636&	0.1482	&0.3969&0.4877	&0.2082&&0.3131&0.3252&	0.1230	&0.4971&0.7494	&0.1625&1.2624\\
	&Bias	      &0.0177&	0.1281&	-0.1012	&0.9610&0.6198 &-0.0920&&0.0029&0.0291&	-0.0414	&0.0726&-0.0281	&0.0301&-0.5433\\
	&Rel.~bias    &0.0707&	0.2562&	-0.5060	&1.6017&-0.3099&-0.3068&&0.0116&0.0582&	-0.2070	&0.1209&0.0141	&0.1004&-0.1358\\
	&MSE	      &0.1456&	0.1486&	0.0322	&1.0811&0.6220	&0.0518&&0.0980&0.1066&	0.0168	&0.2524&0.5624	&0.0273&1.8888\\
\hline
1000&Estimate     &0.2932&	0.5956&	0.1167	&1.5427&-1.3684	&0.2073&&0.2672&0.5125&	0.1634	&0.5705	&-2.1282&0.3320&3.9097\\
	&Std.~error   &0.1249&	0.1330&	0.0377	&0.1406&0.1583	&0.0437&&0.2113&0.2222&	0.0855	&0.3506	&0.5550	&0.1034&1.0105\\
	&SB~std.~err. &0.2694&	0.2644&	0.1139	&0.3152&0.4282	&0.1783&&0.2160&0.2235&	0.0831	&0.3980	&0.7061	&0.1239&1.1502\\
	&Bias	      &0.0432&	0.0956&	-0.0833	&0.9427&0.6316 &-0.0927&&0.0172&0.0125&	-0.0366	&-0.0295&-0.1282&0.0320&-0.0903\\
	&Rel.~bias    &0.1728&	0.1913&	-0.4166	&1.5712&-0.3158&-0.3090&&0.0687&0.0249&	-0.1831	&-0.0491&0.0641	&0.1066&-0.0226\\
	&MSE	      &0.0744&	0.0790&	0.0199	&0.9880&0.5823	&0.0404&&0.0470&0.0501&	0.0082	&0.1593	&0.5150	&0.0164&1.3311\\
\hline \hline
\end{tabular}
}
\end{center}
\end{table}
\end{landscape}

\clearpage

\end{document}